\documentclass[lettersize,journal]{IEEEtran}
\usepackage{amsmath,amsfonts}
\usepackage{algorithmic}
\usepackage{algorithm}
\usepackage{array}
\usepackage[caption=false,font=normalsize,labelfont=sf,textfont=sf]{subfig}
\usepackage{textcomp}
\usepackage{stfloats}
\usepackage{url}
\usepackage{verbatim}
\usepackage{graphicx}
\usepackage{cite}
\usepackage{amssymb}
\usepackage{xcolor}
\newtheorem{Lemma}{Lemma}
\newtheorem{proposition}{Proposition}
\newtheorem{theorem}{Theorem}
\newtheorem{remark}{Remark}

\begin{document}

\title{Intelligent Reflecting Surface-Assisted Symbiotic Radio Systems: A Double-Reflection Covert Communication Design}

\author{Yunpeng Feng, Jian Chen,~\IEEEmembership{Member,~IEEE}, Lu Lv,~\IEEEmembership{Member,~IEEE},  Yuchen Zhou,~\IEEEmembership{Member,~IEEE},\\ Long Yang,~\IEEEmembership{Senior Member,~IEEE}, Naofal Al-Dhahir,~\IEEEmembership{Fellow,~IEEE}, Fumiyuki Adachi,~\IEEEmembership{Life Fellow,~IEEE}

\thanks{Y. Feng, J. Chen, L. Lv, Y. Zhou, and L. Yang are with the State Key Laboratory of Integrated Services Networks, Xidian University, Xi'an 710071, China (e-mail: yunpengfeng2021@163.com; jianchen@mail.xidian.edu.cn;  lulv@xidian.edu.cn; ychenzhou@163.com; lyang@xidian.edu.cn).}

\thanks{Naofal Al-Dhahir is with the Department of Electrical and Computer Engineering, The University of Texas at Dallas, Richardson, TX 75080 USA (e-mail: aldhahir@utdallas.edu).}

\thanks{Fumiyuki Adachi is with the International Research Institute of Disaster Science, Tohoku University, Sendai 980-8577, Japan (e-mail: adachi@ecei.tohoku.ac.jp).}
\thanks{}}


\IEEEpubid{}

\maketitle

\begin{abstract}
We investigate covert communication in an intelligent reflecting surface (IRS)-assisted symbiotic radio (SR) system under the parasitic SR (PSR) and the commensal SR (CSR) cases, where an IRS is exploited to create a double reflection link for legitimate users and degrade the detection performance of the warden (W).
Specifically, we derive an analytical expression for the average detection error probability of W and design an optimal strategy to determine the transmit power and backscatter reflection coefficient. To further enhance the covert performance, the joint optimization of the source transmit power, backscatter device (BD) reflection coefficient, and IRS phase-shifter is formulated as an expectation-based quadratic-fractional (EQF) problem.
By reformulating the original problem into a fraction-eliminated backscatter power leakage minimization problem, we further develop the phase alignment pursuit and the power leakage minimization algorithms for the PSR and the CSR cases, respectively. Numerical results confirm the accuracy of the derived results and the superiority of our proposed strategy in terms of covertness.

\end{abstract}

\begin{IEEEkeywords}
Symbiotic radio, Covert communication, Intelligent reflecting surface. 
\end{IEEEkeywords}

\section{Introduction}

Driven by the ever-increasing number of wireless devices and applications, such as the Internet of Things (IoT),  intelligent transport, telemedicine, and holographic video, the sixth generation (6G) mobile network faces an unprecedented challenge in terms of spectrum resource allocation and energy provisioning {\cite{6G_0,6G_1,6G_2,6G_3}}.
Against this background, symbiotic radio (SR), as a promising communication paradigm for future wireless networks, was proposed as an effective solution to achieve spectral- and energy-efficient wireless communications \cite{SR}.
Specifically, the typical SR communication paradigm usually employs a backscatter device (BD) that modulates its own information onto the continuous primary signals rather than generating its own carrier, thereby enjoying low-power consumption and high energy efficiency {\cite{BD_1,BD_2,BD_3}}.

Due to the great potential for the ubiquitous deployment of SR communication, the soaring volume of confidential and sensitive data (e.g., financial details \cite{financial}, health status monitoring \cite{health}, and identity authentication \cite{identity}) can be transmitted using SR communication.
The communication behavior may leak the state and location information of users \cite{LPD}, which means protecting the confidentiality of communications through information-theoretical security techniques is insufficient in some scenarios \cite{financial,health,identity}, and the communication itself is often required to be hidden from being detected.
Moreover, with the rapid advancements in eavesdropping technologies, the hardware/software-based features of backscattered signals are vulnerable to signal counterfeit, signal relay and replay attacks {\cite{identity,identity2,identity3}}, since eavesdroppers can learn constant signal features or record RF signals to mimic the legitimate BD without completely decoding the signal.
To incorporate stealth capabilities or possess a low probability of detection in SR communication systems, recent advances leverage the signal/spatial degrees of freedom to create the detection uncertainty at the warden (W){\cite{covert_BD1,covert_BD2,covert_BD3,AN_source,AN_destination,AN_jammer}}. 
Specifically, by varying the reflection coefficient of the BD's signal, the authors of \cite{covert_BD1} achieved the upper bound of the covert rate in ambient backscatter systems.
To better hide the covert information, the authors in \cite{covert_BD2} employed the multi-antenna BD to simultaneously modulate the covert and overt signal by the optimal beamforming design.
To further increase the spatial uncertainty, the work in \cite{covert_BD3} leveraged the randomly distributed BDs to produce uncertain aggregate interference.
Moreover, existing works attempt to further degrade the detection performance of W by generating artificial noise at the source (S) \cite{AN_source}, destination \cite{AN_destination}, and additional jammer \cite{AN_jammer}.

As an innovative technology, intelligent reflecting surface (IRS) provides a vital solution to reconfigure the wireless propagation environment which benefits the SR communications {\cite{IRS_SR1,IRS_SR2,IRS_SR3,IRS_SR4}}. 
In fact, IRS can also create additional spatial degrees of freedom, thereby enhancing the uncertainty of the signal detection at W while improving the covert performance of the SR communication systems \cite{IRS_SR5}. 
Considering that the signal received by W only experiences a single reflection from the IRS, the authors in  \cite{semi-passive0,semi-passive1,semi-passive2} investigated the semi-passive IRS as the secondary transmitter that modulates its own information over the incident signal from the primary transmitter.
Specifically, the work \cite{semi-passive0} jointly optimized the phase shifts and reflection amplitude of the semi-passive IRS to align the signal phase at the receiver (R) under the covert constraint. The study in \cite{semi-passive1} analyzed the maximum covert throughput of the SR system and optimized the phase shifts of IRS via the semidefinite relaxation algorithm. To further improve the covert performance, the authors in \cite{semi-passive2} developed a deep unfolding algorithm based on gradient descent for the IRS beamforming design.

As for the aforementioned research efforts, we make two important observations below.
\begin{itemize}
    \item The works in \cite{semi-passive0,semi-passive1,semi-passive2} only integrate the IRS and BD in the parasitic SR (PSR) case. 
    However, in many scenarios, the transmission rate of the BD signal may be much lower than the primary signal. Such a transmission paradigm can be considered as the commensal SR (CSR), which can be applied to smart homes \cite{SR} and identity authentication {\cite{identity,identity2,identity3}}. 
    Theoretically, in the CSR case, the BD signal can be treated as an additional multi-path component rather than interference, which allows R to better decode and eliminate the primary signal. Therefore, how to leverage the characteristics of the CSR transmission paradigm to enhance the covertness of the IRS-assisted BD system is worthy of further investigation.
    \item Existing works in \cite{semi-passive0,semi-passive1,semi-passive2} only consider the semi-passive IRS-assisted covert PSR communication system, where the signals are reflected by the IRS only once to create the uncertainty at W. In fact, when a dedicated BD coexists with the passive IRS, additional uncertainty can be achieved by double signal reflection between the IRS and the BD. However, how to fully utilize the double reflection to improve the covert performance of SR communication systems remains an open issue.
\end{itemize}

Motivated by the above observations, this paper investigates strategy design, performance analysis, and optimization of both PSR and CSR paradigms for covert communications assisted by a passive IRS. The main contributions of this work are listed below:
\begin{itemize}
    \item We propose an IRS-assisted covert communication strategy under both PSR and CSR cases. The integration of the passive IRS and SR makes a double reflection link for the backscatter signal, leading to a fast-fading channel and enhancing the uncertainty of W's signal detection. 
    \item We derive an analytical closed-form expression for the average detection error probability (DEP) of the proposed IRS-assisted covert SR communication strategy. 
    Further, considering the statistical warden channel state information (WCSI) and non-WCSI scenarios, we develop an optimal strategy to determine the transmit power and the backscatter reflection coefficient for each WCSI scenario.
    \item We formulate the average DEP maximization problem as an expectation-based quadratic-fractional (EQF) problem by jointly optimizing the transmit power at S, the reflection coefficient of BD, and the phase-shifter of IRS. 
    By employing the designed strategy, the original problem is reformulated as the fraction-eliminated backscatter power leakage minimization (BPLM) problem, with which we further develop the phase alignment pursuit (PAP) and the power leakage minimization (PLM) algorithm to optimize the phase shifts of IRS for the PSR and CSR cases, respectively.
    \item Through analytical and numerical results, we obtain two useful insights:
    i) When the transmission rate of the primary system is higher than the BD system, the CSR strategy outperforms the PSR strategy in terms of covertness. On the contrary, when the BD system has a certain transmission rate requirement, the PSR strategy has a superior covert performance. ii) The incorporation of a passive IRS creates a double-reflection channel for BD signal detection, thereby increasing the number of IRS elements can enhance covert performance while saving the transmit power.
\end{itemize}

The rest of the paper is organized as follows. Section II introduces the system model of the IRS-assisted SR covert communication. Sections III, IV and V provide the transmission strategy, performance analysis and power/beamforming optimization design for the introduced system, respectively. Numerical results are presented in Section VI. Finally, we conclude the paper in Section VII.

\textit{Notations:} Lowercase and uppercase boldface letters denote vectors and matrices, respectively; $(\cdot)^T$, $(\cdot)^H$, $\text{Tr}(\cdot)$, $\left|\cdot\right|^2$ and $\mathbb{E}\left[\cdot\right]$ symbolize the transpose, conjugate transpose, trace, modulus squared and expectation operations, respectively; $\left\|\boldsymbol{v}\right\|$ denotes the 2-norm of vector $\boldsymbol{v}$; ${\bf I}_M$ represents the $M \times M$ identity matrix; $\left\lceil \cdot \right\rceil$ is the operator of rounding up to the nearest integer.

\begin{figure}
\centering
\includegraphics[width=0.3\textwidth]{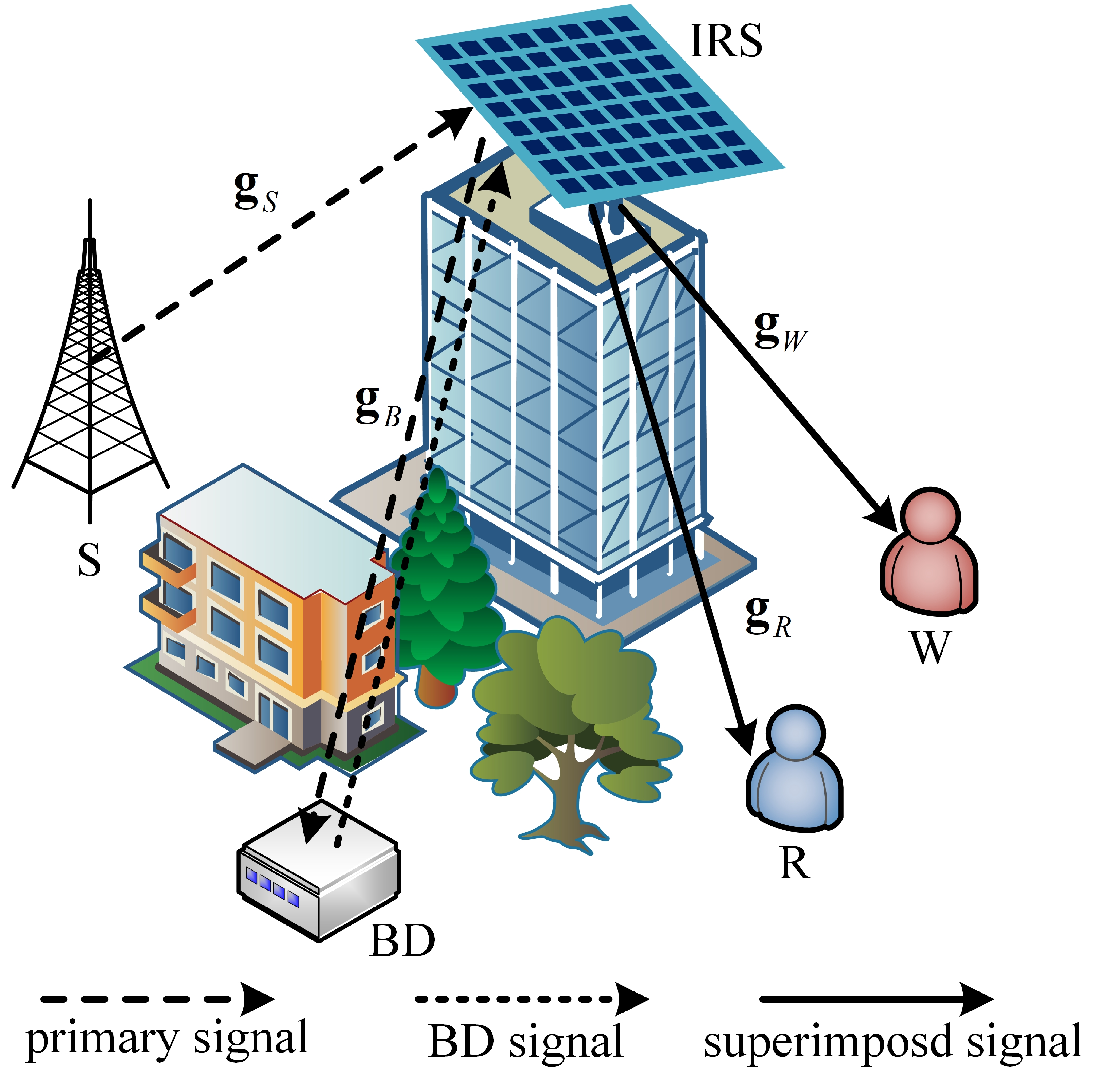}
\caption{{Covert communication in IRS-assisted SR system.}}
\label{fig_sys}
\vspace{-10pt}
\end{figure}

\section{System Model}

Consider that an IRS assists the covert SR communication system, which consists of an S, an IRS, an R, a BD, and a W, as shown in Fig.1.
We assume a special scenario that S, R, BD and W are equipped with a single antenna, and the direct links between S, BD, R and W are blocked by obstacles \cite{YXH}. The IRS creates line-of-sight (LoS) links for all communication nodes through $M$ reconfigurable reflecting elements. S attempts to transmit the primary signal to R using the constant modulus modulation. 
At the same time, BD employs binary phase-shift keying (BPSK) modulation to superimpose its confidential information onto the primary signal to transmit a covert signal. Besides, the IRS is employed to assist the legitimate transmission between S, BD and R, {while improving the uncertainty of W's signal detection}.  
{Considering the severe path loss, we assume that the signals reflected by the IRS two or more times can be ignored \cite{IRS-2,IRS-3}.}

Block fading channels are assumed in this paper, where the channel coefficients remain unchanged within each coherence time but are independent between different fading blocks \cite{blockfading}. During each fading block, the channel coefficient vectors between S/BD/R/W and IRS are denoted by ${\bf g}_S$, ${\bf g}_B$, ${\bf g}_R$ and ${\bf g}_W$, respectively, where $\{{\bf g}_i\}_{i={ S,B,R,W}}\in\mathbb{C}^{M\times 1}$. Consider that $\{{{\bf g}_i}\sim\mathcal{CN}\left(0,1\right)\}_{i={S,B,R,W}}$ are independent and identically distributed (i.i.d.).
For the legitimate links, S can perfectly obtain the instantaneous CSI by IRS-aided orthogonal pilot exchange \cite{pilot}. For the monitoring link, S also knows the statistical WCSI by utilizing the distance information between IRS and W. Considering the worst-case assumption for covert communication, it is assumed that W obtains the perfect knowledge of the instantaneous CSI of the S-IRS-W and BD-IRS-W links. In addition, we assume BD works in a full-duplex mode for signal reception and transmission, thus the channel reciprocity holds.

\vspace{-10pt}
\section{Transmission Strategy and Achievable Rate Analysis}

In this section, we proposed an IRS-assisted covert communication strategy under the PSR and the CSR cases.
{Specifically, in the PSR case, we employ the successive interference cancellation (SIC) technique to remove the interference generated from the primary transmitter. However, in the CSR case, the backscatter signal can be regarded as a multi-path component of the channel rather than interference, which allows R to better decode and eliminate the primary signal.}

For both PSR and CSR cases, we analyze the maximum achievable rate performance. Consider that S transmits the {primary} signal $s(n)(n=1,...,N)$ with the symbol period $T_s$. 
At the same time, BD modulates the covert signal $c(k)(k=0,...,K-1)$ with the symbol period $T_c$.
{According to} the different relationships between $T_s$ and $T_c$, we consider two cases. 1) The PSR case, where the symbol period of the {primary signal equals to the BD signal}, i.e., $T_s=T_c$. 2) The CSR case, where the symbol period of the {primary signal is much less than the BD signal, i.e., $T_c=\eta T_s$, where $\eta\left(\gg1\right)$ is the ratio of $c(k)$ to $s(n)$ in terms of the symbol period.}

\vspace{-10pt}
\subsection{Transmission strategy in the PSR Case}

{In this subsection, we design a transmission strategy in the PSR case for covert communications. With the help of IRS, R receives the primary signal $s(n)$ and the BD signal $c(k)$ generated from S and BD, respectively. Note that the modulation of signal $c(k)$ only employs the passive elements of the BD, thus the backscatter signal $\sqrt{\alpha}c(k)$ has no additive noise, where $\alpha\in[0,1]$ is the power reflection coefficient.}
Assuming that $s(n)$ and $c(k)$ are perfectly synchronized at R, the received signal $y_R^{PSR}(n)$ can be expressed as
\begin{equation}
\begin{array}{ll}
    \label{R_receive_PSR}
    y_R^{PSR}(n)&=\sqrt{\alpha p}\underbrace{\frac{({\bf g}_S^H {\boldsymbol{\Theta}} {\bf g}_B)}{\sqrt{L(d_S) L(d_B)}}}_{\triangleq h_{SB}}\underbrace{\frac{({\bf g}_B^H {\boldsymbol{\Theta}} {\bf g}_R)}{\sqrt{L(d_B)L(d_R)}}}_{\triangleq h_{BR}}s(n)c(k)\\
    &+\sqrt{p}\underbrace{\frac{({\bf g}_S^H {\boldsymbol{\Theta}} {\bf g}_R)}{\sqrt{L(d_S)L(d_R)}}}_{\triangleq h_{SR}}s(n)+n_R(n),
\end{array}
\end{equation}
where $p$ is the transmit power of S, $L(d)$ represents the effective path loss function, ${h}_{ij}$ is the cascade channel of the $i$-IRS-$j$ link for $i,j\in\{{S,B,R}\}$, 
${\boldsymbol{\Theta} = \text{diag}({{e^{j{\theta _1}}},{e^{j{\theta _2}}},...,{e^{j{\theta _M}}}})}$ denotes the IRS phase-shift matrix for {${\theta_m\in[0,2\pi]}$} and $n_R\sim\mathcal{CN}(0,\sigma_0^2)$ represents the additive noise at R.

It is worth noting that the BD signal $c(k)$ suffers from double channel fading, i.e., $h_{SB}$ and $h_{BR}$. Thus, the decoding order of R follows $s(n)\to c(k)$ when performing the SIC technique. First, for the detection of $s(n)$, the received signal component $\sqrt{\alpha p} {h_{SB}} {h_{BR}}{s(n)}{c(k)}$ can be treated as interference. Since $\mathbb{E}\left[\left|s(n)\right|^2\right]=\mathbb{E}\left[\left|c(k)\right|^2\right]=1$, the average interference power can be represented as $\mathbb{E} \left[ \alpha p \left|{h_{SB}}\right|^2 \left|{h_{BR}}\right|^2 \left|{s(n)}\right|^2 \left|{c(k)}\right|^2 \right] = \alpha p \left|{h_{SB}}\right|^2 \left|{h_{BR}}\right|^2$. Therefore, the maximum achievable data rate of R to decode $s(n)$ is given by \footnote{In this work, we consider that the capacity-achieving coding scheme is employed. Thus, when the actual data rate is no more than the maximum achievable data rate, says Shannon capacity, there always exists a coding scheme to ensure the successful decoding with arbitrary small bit error probability \cite{Shannon}.}
\begin{equation}
    \label{PSR_rate_s}
    R_s^{PSR}=\log_2\left(1+\frac{p\left|{h_{SR}}\right|^2}{\alpha p \left|{h_{SB}}\right|^2 \left|{h_{BR}}\right|^2 +\sigma_0^2}\right).
\end{equation}

Note that the interference is the multiplication of two complex Gaussian channels $h_{SB}$ and $h_{BR}$, and it thus follows a non-Gaussian distribution \cite{SR}. Before decoding $c(k)$, R employs the SIC technique to remove $s(n)$. It can be observed that $s(n)$ generates the fast fading channel $h_{fast}=\sqrt{\alpha p}{h_{SB}}{h_{BR}}s(n)$ over $c(k)$. 
Consider the block fading channel model where the ${h_{SB}}{h_{BR}}$ is fixed in each channel coherence time, and the variation of $h_{fast}$ is induced by $s(n)$.
Thus, by taking a mathematical expectation over $s(n)$, the maximum achievable data rate of $c(k)$ under the fast-fading channel can be given by
\begin{equation}
    \label{PSR_rate_c}
    R_c^{PSR}=\mathbb{E}_s \left[ \log_2\left(1+\frac{\left|h_{fast}\right|^2}{\sigma_0^2}\right) \right].
\end{equation}

\begin{figure}
\centering
\includegraphics[width=0.4\textwidth]{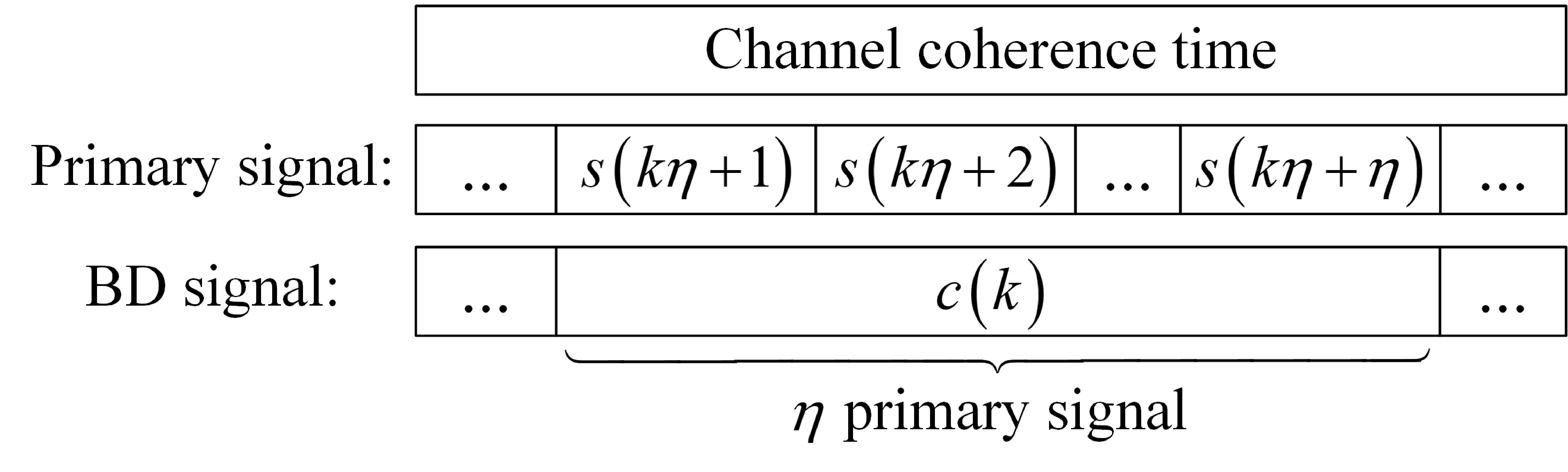}
\caption{{Transmission frame for the CSR case within a channel coherence time.}}
\label{frame}
\vspace{-10pt}
\end{figure}

\vspace{-10pt}

\subsection{Transmission strategy in the CSR Case}

As shown in Fig. \ref{frame}, the CSR case assumes that the signal $c(k)$ covers $\eta$ symbol periods of the signal $s(n)$, i.e., $N=\eta K$, $T_c=\eta T_s$, $k=\left\lceil {\frac{n}{\eta}} \right\rceil -1$. Therefore, in the $n$-th symbol period of the signal $s(n)$, the received signal $y_R^{CSR}$ can be written as
\begin{equation}
    \label{R_receive_CSR}
    y_R^{CSR}(n)=\sqrt{p}{ h_{SR}}s(n)+\sqrt{\alpha p}{ h_{SB}}{ h_{BR}}\cdot c(k)\cdot s(n)+n_R(n).
\end{equation}

Since the symbol period of the BD signal $c(k)$ is much longer than the primary signal $s(n)$, $c(k)$ can be considered as a multi-path component of the channel rather than interference when R decodes $s(n)$, which means that $c(k)$ is invariant as a constant during the symbol period of $s(n)$. Thus, the equivalent channel for decoding $s(n)$ can be represented as ${ h}_{ SR}^{CSR}={ h_{SR}}+\sqrt{\alpha}{ h_{SB}}{ h_{BR}}c(k)$. 
Since $\eta$ is assumed to be large enough \cite{coherence}, the maximum achievable data rate of $s(n)$ can be approximated as
\begin{equation}
    \label{CSR_rate_s}
    R_s^{CSR}=\mathbb{E}_c\left[ \log_2\left(1+\frac{p\left|{ h}_{ SR}^{CSR}\right|^2}{\sigma_0^2}\right)\right].
\end{equation}

Assume that during a symbol period of $c(k)$, $s(n)(n=k\eta +1,...,k\eta +\eta )$ can be successfully decoded, and R employs the SIC technique to remove the interference caused by the primary signal to the BD signal.
Considering the perfect decoding and cancellation of $s(n)$, the remaining received signal at the $k$-th symbol period of $c(k)$ can be expressed as follows
\begin{equation}
    \label{R_receive_CSR_remaining}
    \tilde{\bf y}_R^{CSR}(k)=\sqrt{\alpha p}{ h_{SB}}{ h_{BR}}\cdot c(k)\cdot {\bf s}(k)+{\bf n}_R(k).
\end{equation}
In this situation, $\tilde{\bf y}_R^{CSR}(k)$, ${\bf s}(k)$ and ${\bf n}_R(k)$ are vectors of length $\eta $, which can be represented as $\tilde{\bf y}_R^{CSR}(k)=\left[\tilde{y}_R^{CSR}(k\eta +1),...,\tilde{y}_R^{CSR}(k\eta +\eta )\right]$, ${\bf s}(k)=\left[s(k\eta +1),...,s(k\eta +\eta )\right]$ and ${\bf n}_R(k)=\left[n_R(k\eta +1),...,n_R(k\eta +\eta )\right]$, respectively. Assuming the recovery of the signal $s(n)$ has been successfully accomplished with $\mathbb{E}\left[\left|s(n)\right|^2\right]=1$, {the maximal ratio combining (MRC) strategy can be employed to increase the signal-to-noise ratio (SNR) of $c(k)$'s decoding.
In this situation, when $\eta $ symbols of $s(n)$ are transmitted, BD only conveys one $c(k)$ symbol. Thus, the maximum achievable data rate of $c(k)$ is decreased by $1/\eta $, which can be expressed as follows.}
\begin{equation}
    \label{CSR_rate_c}
    R_c^{CSR}=\frac{1}{\eta }\log_2\left(1+\frac{\eta \alpha p \left|{ h_{SB}}\right|^2 \left|{ h_{BR}}\right|^2 }{\sigma_0^2}\right).
\end{equation}

According to \eqref{PSR_rate_s}, \eqref{PSR_rate_c}, \eqref{CSR_rate_s} and \eqref{CSR_rate_c}, it is worth noting that in the PSR and the CSR cases, the transmission performance is primarily affected by the rate requirement of $s(n)$ and $c(k)$, respectively. Moreover, the ratio of the symbol period between the primary and the BD signal greatly limits the rate requirement of $c(k)$.

\subsection{Covert Signal Detection at Warden}

{To decide whether BD generates the covert signal or not, W faces a binary detection problem:} 1) the null hypothesis $\mathcal{H}_0$ where BD does not transmit $c(k)$ to R; 2) the alternative hypothesis $\mathcal{H}_1$ where BD has transmitted $c(k)$ to R. Then, considering the PSR and the CSR cases, the received signals at W under $\mathcal{H}_0$ and $\mathcal{H}_1$ are, respectively, given by

\subsubsection{The PSR case}
\begin{equation}
\begin{array}{ll}
    \label{H_0_PSR}
    {\mathcal H_0}: &y_W^{PSR}(n)=\sqrt{p}\underbrace{\frac{({\bf g}_S^H {\boldsymbol{\Theta}} {\bf g}_W)}{\sqrt{L(d_S)L(d_W)}}}_{\triangleq h_{SW}}s(n)+n_W(n),\\
\end{array}
\end{equation}
\begin{equation}
\begin{array}{ll}
    \label{H_1_PSR}
    {\mathcal H_1}: &y_W^{PSR}(n)=\sqrt{\alpha p}{ h_{SB}}\underbrace{\frac{({\bf g}_B^H {\boldsymbol{\Theta}} {\bf g}_W)}{\sqrt{L(d_B)L(d_W)}}}_{\triangleq  h_{BW}}s(n)c(k)\\
    &+\sqrt{p}{ h_{SW}}s(n)+n_W(n),
\end{array}
\end{equation}
where $n_W \sim \mathcal{CN}\left(0,\sigma_0^2\right)$ represents the additive noise at W.

\subsubsection{The CSR case}

\begin{equation}
\begin{array}{ll}
    \label{H_0_CSR}
    {\mathcal H_0}:&{\bf y}_W^{CSR}(k)=\sqrt{p}{ h_{SW}}{\bf s}(k)+{\bf n}_W(k),
\end{array}
\end{equation}
\begin{equation}
\begin{array}{ll}
    \label{H_1_CSR}
    {\mathcal H_1}:&{\bf y}_W^{CSR}(k)=\sqrt{\alpha p}{ h_{SB}}{ h_{BW}}{\bf s}(k)c(k)\\
    &+\sqrt{p}{ h_{SW}}{\bf s}(k)+{\bf n}_W(k),
\end{array}
\end{equation}
where ${\bf y}_W^{CSR}(k)=\left[y_W^{CSR}(k\eta +1),...,y_W^{CSR}(k\eta +\eta )\right]$, ${\bf s}(k)=\left[s(k\eta +1),..,s(k\eta +\eta )\right]$ and ${\bf n}_W(k)=\left[n_W(k\eta +1),...,n_W(k\eta +\eta )\right]$ are vectors of length $\eta$. 

{According to the Neyman-Pearson criterion \cite{Hypotheses}, the decision rule of W is given by}

\begin{equation}
    P_{W}\mathop{\gtrless}\limits_{\mathcal{D}_{0}}^{\mathcal{D}_{1}} \tau,
\end{equation}
where $P_W$ is the average received power at W, $\tau(>0)$ is the detection threshold of W, ${\mathcal D}_1$ and ${\mathcal D}_0$ are binary decisions in favor of ${\mathcal H}_1$ and ${\mathcal H}_0$, respectively. 
Specifically, for the PSR and CSR cases, $P_W$ is defined as $P_W^{PSR}\triangleq\frac{1}{N}\sum\limits_{n=1}^N \left|y_W^{PSR}(n)\right|^2$ and $P_W^{CSR}\triangleq\frac{1}{N}\sum\limits_{k=0}^{K-1} \left\|{\bf y}_W^{CSR}(k)\right\|^2$, respectively.
Assume that $N$ and $K$ are sufficiently large. Thus, the average received power expression of W is identical for both the PSR and CSR cases, which can be represented as follows
\begin{equation}
\label{P_W}
P_{W}=\left\{\begin{array}{ll}
p \left|{ h_{SW}}\right|^2 +\sigma_0^2, & \mathcal{H}_{0}, \\
\alpha p \left|{ h_{SB}}\right|^2 \left|{ h_{BW}}\right|^2 + p \left|{ h_{SW}}\right|^2 +\sigma_0^2, &\mathcal{H}_{1}.
\end{array}\right.
\end{equation}

{Using the given $\mathcal{H}_i$ and $\mathcal{D}_i$ for $i\in\{0,1\}$, the false alarm and miss detection probabilities of W can be defined as ${\mathbb P}_{\text {FA}}\triangleq{\mathbb P}({\mathcal D}_1|{\mathcal H}_0)$ and ${\mathbb P}_\text{MD}\triangleq{\mathbb P}({\mathcal D}_0|{\mathcal H}_1)$, respectively. 
Then, we evaluate the performance of W's hypothesis test by the average DEP, which is defined as}
\begin{equation}
\label{error_Pro}
\xi_{W,\tau}\triangleq {\mathbb P}_{\text{FA}} + {\mathbb P}_{\text{MD}},
\end{equation}
where $0 \leq \xi_{W,\tau} \leq 1$.

\begin{remark}
    As can be observed from \eqref{PSR_rate_c} and \eqref{CSR_rate_s}, $s(n)$ and $c(k)$ provide additional uncertainty from modulated symbols, which also confuses the signal detection at W. Additionally, the integration of the IRS and SR introduces a double reflection link for $c(k)$, which creates a fast-fading channel ${ h_{SB}h_{BW}}s(n)$ and further increases the uncertainty of W’s signal detection.
\end{remark}

\section{Covert Performance Anaysis}

{In this section, we analyze the covert performance of the SR communication system assisted by a passive IRS. The average DEP of the proposed transmission strategy is derived in closed form. Then, considering the observability of W’s CSI, we developed an optimal strategy to determine the source transmit power and the backscatter reflection coefficient from the derived results.}

To derive the average DEP of W, we approximated the CDF of $\left({\bf g}_S^H {\bf \Theta} {\bf g}_W\right)$, $\left({\bf g}_S^H {\bf \Theta g}_B\right)$ and $\left({\bf g}_B^H {\bf \Theta g}_W\right)$
as a complex Gaussian random variable $\mathcal{K}\sim\left(0,M\right)$ \cite{tight_M}.
Therefore, the false alarm probability of W is derived as  
\begin{equation}
\label{Pro_FA}
\begin{array}{ll}
\mathbb{P}_{\text{FA}} & ={\text{Pr}}\left(\frac{p\left|{\bf g}_S^H {\bf \Theta} {\bf g}_W \right|^2}{L(d_S) L(d_W)}+\sigma_0^2 > \tau \right)\\
& = \left\{
\begin{array}{ll}
     1, & \tau \leq \sigma_0^2, \\
     e^{-\lambda l_1 z},  & \tau > \sigma_0^2,
\end{array}
\right.
\end{array}
\end{equation}
where $l_1=L(d_S) L(d_W)$, $\lambda=\frac{1}{M}$ and $z=\frac{\tau-\sigma_0^2}{p}$. 
Subsequently, we derive the miss detection probability of W with the help of the following proposition.
\begin{proposition}
    For the case of $M\to\infty$, the complex Gaussian random variables ${\bf g}_S^H {\bf \Theta g}_B$ and ${\bf g}_B^H {\bf \Theta g}_W$ are i.i.d.    
\end{proposition}
\begin{IEEEproof}
    Please refer to Appendix \ref{App_A}.
\end{IEEEproof}

{Thus, the PDF of $\left|{\bf g}_S^H {\bf \Theta} {\bf g}_B \right|^2\times\left|{\bf g}_B^H {\bf \Theta} {\bf g}_W \right|^2$ is obtained from Appendix \ref{App_A} in \cite{exp_PDF}. We employ the Gaussian-Chebyshev quadrature \cite{G-C} to approximate the miss detection probability of W as follows.}
\begin{equation}
\label{Pro_MD}
\begin{array}{ll}
&\mathbb{P}_{\text{MD}}={\text{Pr}}\left(
\frac{\alpha p\left|{\bf g}_S^H {\bf \Theta} {\bf g}_B \right|^2}{L(d_S) L(d_B)}
\frac{\left|{\bf g}_B^H {\bf \Theta} {\bf g}_W \right|^2}{L(d_B) L(d_W)}+\frac{p\left|{\bf g}_S^H {\bf \Theta} {\bf g}_W \right|^2}{L(d_S) L(d_W)}\right.\\
& \quad\quad\quad\quad\quad\quad +\left. \sigma_0^2 < \tau 
\right)\\
& = \left\{
\begin{array}{ll}
     \quad\quad\quad 0, & \tau \leq \sigma_0^2, \\
     \begin{array}{ll}
         & 2\lambda^2 \int_{0}^{\frac{l_2 z}{\alpha}} \left(1-e^{-l_1 \lambda z+\frac{l_1}{l_2}\alpha \lambda x}\right) \\
         & \times K_0\left(2 \lambda \sqrt{x} \right) {\text{dx}},   
     \end{array}
     & \tau > \sigma_0^2,
\end{array}
\right.
\end{array}
\end{equation}
\begin{equation}
\begin{array}{ll}\label{MD_G_C}
     & = \left\{
     \begin{array}{ll}
          \quad 0, & \tau \leq \sigma_0^2,  \\
          \begin{array}{ll}
        1-2 \lambda \sqrt{\frac{l_{2} z}{\alpha}} K_{1}\left(2 \lambda \sqrt{\frac{l_{2} z}{\alpha}}\right)\\
        -\frac{\lambda^{2} l_{2} z \pi}{\alpha Q} \sum\limits_{q=1}^{Q}\left\{\sqrt{1-x_{q}^{2}} \cdot e^{\frac{\lambda l_{1} z\left(x_{q}-1\right)}{2}} \right.\\
        \left.\times K_{0}\left(\lambda \sqrt{\frac{2 l_{2} z\left(x_{q}+1\right)}{\alpha}}\right)\right\},
          \end{array}
            & \tau > \sigma_0^2,\\
     \end{array}
     \right.
\end{array}
\end{equation}
where $l_2=l_1 L(d_B)^2$, $K_0(x)$ is the modified Bessel function of the second kind and $Q$ is a complexity-accuracy tradeoff parameter. Substituting \eqref{Pro_FA} and \eqref{MD_G_C} into \eqref{error_Pro}, the approximated closed-form expression of W's average DEP can be obtained as
\begin{equation}
\begin{array}{ll}\label{DEP}
     &\xi_{W,\tau}  \\
     & = \left\{
     \begin{array}{ll}
          \quad 1, & \tau \leq \sigma_0^2,  \\
          \begin{array}{ll}
            2\lambda^2 \int_{0}^{\frac{l_2 z}{\alpha}} \left(1-e^{-l_1 \lambda z+\frac{l_1}{l_2}\alpha \lambda x}\right)  \\
            \times K_0\left(2 \lambda \sqrt{x} \right) {\text{dx}} + e^{-\lambda l_1 z}, 
          \end{array}
            & \tau > \sigma_0^2,\\
     \end{array}
     \right.
\end{array}
\end{equation}
\begin{equation}
\begin{array}{ll}\label{DEP_G_C}
     & = \left\{
     \begin{array}{ll}
          \quad 1, & \tau \leq \sigma_0^2,  \\
          \begin{array}{ll}
        e^{-\lambda l_{1} z}+1-2 \lambda \sqrt{\frac{l_{2} z}{\alpha}} K_{1}\left(2 \lambda \sqrt{\frac{l_{2} z}{\alpha}}\right)\\
        -\frac{\lambda^{2} l_{2} z \pi}{\alpha Q} \sum\limits_{q=1}^{Q}\left\{\sqrt{1-x_{q}^{2}} \cdot e^{\frac{\lambda l_{1} z\left(x_{q}-1\right)}{2}} \right.\\
        \left.\times K_{0}\left(\lambda \sqrt{\frac{2 l_{2} z\left(x_{q}+1\right)}{\alpha}}\right)\right\},
          \end{array}
            & \tau > \sigma_0^2.\\
     \end{array}
     \right.
\end{array}
\end{equation}

Note that the average DEP in \eqref{DEP_G_C} is related to $\tau$, $p$ and $\alpha$.
Considering the observability of statistical CSI at W, we first present the optimal solution of $\tau$ in the following theorem.
\begin{theorem}
    {The optimal detection threshold of W for minimizing the average DEP is obtained by numerical integration that satisfies the following equation:} 
    \begin{equation}
        \int_{0}^{\frac{l_2 (\tau-\sigma_0^2)}{\alpha p}} e^{\frac{l_1 \alpha \lambda x}{l_2}} K_0\left(2 \lambda \sqrt{x} \right) {\text{dx}} = \frac{1}{2\lambda^2}.
    \end{equation}
\end{theorem}
\begin{IEEEproof}
    Please refer to Appendix \ref{App_B}.
\end{IEEEproof}

However, if W is a completely passive node, the statistical WCSI is unobservable by the legitimate system, which means that we cannot obtain the optimal detection threshold of W. In this case, we assume that W adopts an arbitrary and fixed detection threshold. 

Overall, with respect to the different detection thresholds of W, we design an optimal strategy to determine the backscatter reflection coefficient $\alpha$ and the transmit power $p$ in the following propositions.

\begin{proposition}\label{Pro2}
    For the optimal or arbitrary detection threshold, the average DEP $\xi_{W,\tau}$ is monotonically decreasing with respect to $\alpha$.
\end{proposition}
\begin{IEEEproof}
    Please refer to Appendix \ref{App_C}.
\end{IEEEproof}
\begin{proposition}\label{Pro3}
    If optimal detection threshold $\tau^\star$ is employed at W, the value of $p$ has no impact on the average DEP. However, when W adopts an arbitrary and fixed detection threshold, the optimal $p$ can be determined at the boundary of its feasible region.
\end{proposition}
\begin{IEEEproof}
    Please refer to Appendix \ref{App_D}.
\end{IEEEproof}

Proposition \ref{Pro3} also reveals that increasing the power of the public signal does not always enhance the covert performance of the proposed SR communication system. This is because it also increases the leakage of covert signal power.

\section{Joint Power and IRS Beamforming Design}

This section investigates the joint transmit power and IRS beamforming design to further enhance the covertness of IRS-assisted SR systems. 
Specifically, we first consider the statistical WCSI scenario and propose the PAP algorithm and the PLM algorithm for the PSR and CSR cases, respectively. Then, the extension of the proposed algorithms to the non-WCSI scenario and the convergence/complexity analysis of the proposed algorithms are discussed at the end of this section.

\subsection{Problem Formulation and Reformulation}

According to the approximated closed-form expression of $\xi_{W,\tau}$ derived in the previous section, we formulate the average DEP maximization problem as an EQF problem in the PSR and CSR cases while ensuring the SIC condition of the primary signal and the quality of service (QoS) constraint of the BD signal.

\subsubsection{The PSR case}

Recall that in the PSR case, the symbol period of the primary signal is equal to the BD signal, 
thus the optimization problem can be formulated as an EQF problem under the results given in \eqref{PSR_rate_s} and \eqref{PSR_rate_c}.
\begin{equation}\nonumber
{\text{P}_1^{PSR}}:\mathop {\max }\limits_{\{ {\mathbf{\Theta}},{\alpha},p \} }  \xi_{W,\tau}\left(\alpha,p\right)
\end{equation}
\begin{equation}
\label{C_power_PSR}
\begin{array}{lc}
\text{s.t.}&p\leq P_{\max}, \,\, 0<\alpha\leq 1,
\end{array}    
\end{equation}
\begin{equation}
\label{C_PSR_diag}
{\mathbf{\Theta}}=\text{diag}\left(e^{j\theta_1},e^{j\theta_2},...,e^{j\theta_M}\right),\, 0\leq \theta_m \leq 2\pi,\, m=1,...,M,
\end{equation}
\begin{equation}
\label{C_Rs_psr}
\log_2\left(1+\frac{p\left|{h_{SR}({\bf\Theta})}\right|^2}{\alpha p \left|{h_{SB}({\bf\Theta})}\right|^2 \left|{h_{BR}({\bf\Theta})}\right|^2 +\sigma_0^2}\right) \geq \epsilon_{SIC},
\end{equation}
\begin{equation}
\label{C_Rc_psr}
\mathbb{E}_s \left[ \log_2\left(1+\frac{\alpha p \left|{h_{SB}({\bf\Theta})}\right|^2 \left|{h_{BR}({\bf\Theta})}\right|^2 \left|{s(n)}\right|^2}{\sigma_0^2}\right) \right]\geq\epsilon_c,
\end{equation}
where \eqref{C_power_PSR} accounts for the transmit power constraint of S and the power reflection constraint of BD, \eqref{C_PSR_diag} denotes the diagonal constant-modulus constraint for the IRS,
{\eqref{C_Rs_psr} guarantees the successful implementation of SIC for the signal $s(n)$, and \eqref{C_Rc_psr} represents the QoS constraint with $\epsilon_c$ being defined as the minimum rate requirement of the BD signal.}
Note that problem ${\text{P}_1^{PSR}}$ is mathematically intractable, due to the complicated expression in the objective function, the non-convex constraint \eqref{C_PSR_diag}, the highly coupled optimization variables $p$, $\alpha$ and ${\mathbf{\Theta}}$ in \eqref{C_Rs_psr} and \eqref{C_Rc_psr}, and the expectation operation of $s(n)$ in \eqref{C_Rc_psr}. To tackle this intractable EQF problem, in the following, we reformulate the original problem with the help of the developed strategy in Propositions 2 and 3.

Since the signal $s(n)$ adopts the constant modulus modulation, the modulus squared of $s(n)$ is always equal to 1. Thus, the QoS constraint of $c(k)$ can be approximated by
\begin{equation}
    \label{R_c_PSR}
    {\mathcal{R}}_c^{PSR}=\log_2\left(1+\frac{\alpha p \left|{h_{SB}({\bf\Theta})}\right|^2 \left|{ h_{BR}({\bf\Theta})}\right|^2}{\sigma_0^2}\right) \geq \epsilon_c.
\end{equation}

According to \eqref{C_Rs_psr} and \eqref{R_c_PSR}, the feasible region of $\alpha$ with a given $\bf \Theta$ can be represented as
\begin{equation}
    \label{alpha_domain}
    \underbrace{\frac{\gamma_c \sigma_0^2}{p \left|{ h_{SB}({\bf\Theta})}\right|^2 \left|{h_{BR}({\bf\Theta})}\right|^2}}_{\triangleq \tilde{\alpha}_{lower}^{PSR}({\bf \Theta},p)} \leq \alpha \leq \underbrace{\frac{p \left|{h_{SR}({\bf\Theta})}\right|^2-{\sigma_0^2} \gamma_{SIC}}{p \gamma_{SIC} \left|{h_{SB}({\bf\Theta})}\right|^2 \left|{h_{BR}(\bf\Theta)}\right|^2}}_{\triangleq \tilde{\alpha}_{upper}^{PSR}({\bf \Theta},p)},
\end{equation}
where $\gamma_c = 2^{\epsilon_c}-1$ and $\gamma_{SIC} = 2^{\epsilon_{SIC}}-1$.
It is worth noting that if ${\tilde{\alpha}_{lower}^{PSR}({\bf \Theta},p)} > 1$ for any value of $\bf \Theta$ and $p$, indicating the nonexistence of a feasible domain for the problem ${\text{P}_1^{PSR}}$.
Consider that $0 <{\tilde{\alpha}_{lower}^{PSR}({\bf \Theta},p)} \leq 1$, the feasible region of $\alpha$ can be obtained as
\begin{equation}
    \label{alpha_region}
    \alpha \in \left[ {\tilde{\alpha}_{lower}^{PSR}({\bf \Theta},p)},\min\left\{1,{\tilde{\alpha}_{upper}^{PSR}({\bf \Theta},p)}\right\}\right].
\end{equation}

As known from Proposition \ref{Pro2}, the average DEP monotonically decreases with $\alpha$, indicating that $\alpha$ should be minimized for problem ${\text{P}_1^{PSR}}$. With the result in \eqref{alpha_region}, the optimal $\alpha$ can be obtained using the lower bound of the feasible region, i.e., $\alpha_{\min}={\tilde{\alpha}_{lower}^{PSR}({\bf \Theta},p)}$.

Besides, for the statistical WCSI scenario, the optimal detection threshold of W can be employed in ${\text{P}_1^{PSR}}$. As mentioned in Proposition \ref{Pro3}, the value of $p$ has no impact on the average DEP under the optimal detection threshold, and as $p$ increases, the feasible domain of $\text{P}_1^{PSR}$ also expands. Therefore, the transmit power of ${\text{P}_1^{PSR}}$ can be determined as $P_{\max}$. For ease of representation, we define $\hat{p}\triangleq P_{\max}$, and the problem $\text{P}_1^{PSR}$ can be reformulated as a BPLM problem.
\begin{equation}\nonumber
{\text{P}_2^{PSR}}:\mathop {\min }\limits_{\mathbf{\Theta}}\quad {\tilde{\alpha}_{lower}^{PSR}({\bf \Theta},\hat{p})}
\end{equation}
\begin{equation}\nonumber
\text{s.t.}\quad\eqref{C_PSR_diag},
\end{equation}
\begin{equation}\label{PSR_C1}
{\tilde{\alpha}_{lower}^{PSR}({\bf \Theta},\hat{p})} \leq {\tilde{\alpha}_{upper}^{PSR}({\bf \Theta},\hat{p})},
\end{equation}
\begin{equation}\label{PSR_C2}
{\tilde{\alpha}_{lower}^{PSR}({\bf \Theta},\hat{p})} \leq 1.
\end{equation}
 
\subsubsection{The CSR case}

In the CSR case, the symbol period of the primary signal is much smaller than the BD signal.
With the result in \eqref{CSR_rate_s} and \eqref{CSR_rate_c}, the optimization problem can also be formulated as an EQF problem.
\begin{equation}\nonumber
{\text{P}_1^{CSR}}:\mathop {\max }\limits_{\{ {\mathbf{\Theta}},{\alpha},p \} }  \xi_{W,\tau}\left(\alpha,p\right)
\end{equation}
\begin{equation}\nonumber
\text{s.t.} \quad \eqref{C_power_PSR},\,\, \eqref{C_PSR_diag},
\end{equation}
\begin{equation}\small
    \label{C_Rs_csr}
    \mathbb{E}_c\left[ \log_2\left(1+\frac{p\left|{h_{SR}({\bf\Theta})}+\sqrt{\alpha}{ h_{SB}({\bf\Theta})h_{BR}({\bf\Theta})}c(k)\right|^2}{\sigma_0^2}\right)\right] \geq \epsilon_{SIC},
\end{equation}
\begin{equation}
    \label{C_Rc_csr}
    \frac{1}{\eta }\log_2\left(1+\frac{\eta \alpha p \left|{h_{SB}({\bf\Theta})}\right|^2 \left|{h_{BR}({\bf\Theta})}\right|^2 }{\sigma_0^2}\right) \geq \epsilon_{c}.
\end{equation}

Similarly, the problem $\text{P}_1^{CSR}$ is also intractable, due to the non-convex objective function, the variable coupling in \eqref{C_Rs_csr} and \eqref{C_Rc_csr}, and the expectation operation in \eqref{C_Rs_csr}. Following the rationale from $\text{P}_1^{PSR}$ to $\text{P}_2^{PSR}$, we can also reformulate $\text{P}_1^{CSR}$.

Recall that signal $c(k)$ employs BPSK modulation, we assume the probability of the signal $c(k)$ being equal to $+1$ or $-1$ is $\frac{1}{2}$. Thus, by generating a large number of realizations of $c(k)$, the constraint \eqref{C_Rs_csr} can be approximated by 
\begin{equation}
    \label{R_s_csr}
    \begin{array}{ll}
         &\frac{1}{2}\log_2\left(1+\frac{p\left|{h_{SR}({\bf\Theta})}+\sqrt{\alpha}{h_{SB}({\bf\Theta})h_{BR}({\bf\Theta})}\right|^2}{\sigma_0^2}\right) \\
         & +\frac{1}{2}\log_2\left(1+\frac{p\left|{h_{SR}({\bf\Theta})}-\sqrt{\alpha}{h_{SB}({\bf \Theta})h_{BR}({\bf\Theta})}\right|^2}{\sigma_0^2}\right) \geq \epsilon_{SIC}.
    \end{array}     
\end{equation}

With some algebraic manipulation, \eqref{R_s_csr} can be re-expressed as follows

\begin{equation}
    \label{re-R_s_csr}
    \begin{array}{cc}
         &\left( \alpha p \left|{h_{SB}({\bf\Theta})}\right|^2 \left|{h_{BR}({\bf\Theta})}\right|^2 + \sigma_0^4 -p \left|{h_{SR}({\bf\Theta})}\right|^2 \right)^2 \\
         & \geq \sigma_0^4\left( 1+\gamma_{SIC} \right)-4 \sigma_0^2 p \left|{h_{SR}({\bf\Theta})}\right|^2.
    \end{array}
\end{equation}

Observe that when $p\left|{ h_{SR}({\bf\Theta})}\right|^2/{\sigma_0^2}\geq\frac{\gamma_{SIC}+1}{4}$, the SIC constraint \eqref{R_s_csr} is always satisfied. However, when $p\left|{ h_{SR}({\bf\Theta})}\right|^2/{\sigma_0^2}<\frac{\gamma_{SIC}+1}{4}$, to satisfy the constraint \eqref{R_s_csr}, the lower bound for $\alpha$ can be obtained according to \eqref{re-R_s_csr}.
\begin{equation}\small
    \label{alpha_range_csr_sic}
    \alpha \geq  \underbrace{\frac{\sqrt{\sigma_0^4\left( 1+\gamma_{SIC} \right)-4 \sigma_0^2 p \left|{ h_{SR}({\bf\Theta})}\right|^2}-\sigma_0^2+3p\left|{ h_{SR}({\bf\Theta})}\right|^2}{p \left|{ h_{SB}({\bf\Theta})}\right|^2 \left|{ h_{BR}({\bf\Theta})}\right|^2}}_{\triangleq \alpha_{lower,SIC}}.
\end{equation}

Following the manipulation from \eqref{R_c_PSR} to \eqref{alpha_domain}, the QoS constraint \eqref{C_Rc_csr} can be re-expressed as 
\begin{equation}
    \label{alpha_range_csr_QoS}
    \alpha \geq \underbrace{\frac{\left(2^{\eta \epsilon_c}-1\right)\sigma_0^2}{p\eta  \left|{ h_{SB}({\bf\Theta})}\right|^2 \left|{ h_{BR}({\bf\Theta})}\right|^2 }}_{\triangleq \alpha_{lower,c}}.
\end{equation}

Recall that to maximize $\xi_{W,\tau}$, we should minimize $\alpha$ and maximize $p$, i.e., $p^\star=\hat{p}$. Based on the above results, we define the SNR of the primary signal as $\text{SNR}_p \triangleq \hat{p} \left|{ h_{SR}({\bf\Theta})}\right|^2$$/{\sigma_0^2}$, and ${\text{P}_1^{CSR}}$ can be reformulated as a BPLM problem under the following two regimes of $\text{SNR}_p$:

\begin{itemize}
    \item Low $\text{SNR}_p$ regime, i.e., $\hat{p} \left|{ h_{SR}({\bf\Theta})}\right|^2/{\sigma_0^2}<\frac{\gamma_{SIC}+1}{4}$, the problem ${\text{P}_1^{CSR}}$ can be reformulated as follows.
\end{itemize}
\vspace{-2pt}
\begin{equation}\nonumber
{\text{P}_2^{CSR,low}}:\mathop {\min }\limits_{\alpha,\mathbf{\Theta}}\quad {{\alpha}}
\end{equation}
\begin{equation}\nonumber
\text{s.t.}\quad\eqref{C_PSR_diag},\, \eqref{alpha_range_csr_sic},\,\eqref{alpha_range_csr_QoS},
\end{equation}
\begin{equation}
0<\alpha\leq 1,
\end{equation}
\begin{equation}\label{39}
    \hat{p} \left|{ h_{SR}({\bf\Theta})}\right|^2/{\sigma_0^2}<\frac{\gamma_{SIC}+1}{4}.
\end{equation}

To tackle the complicated fractional terms in \eqref{alpha_range_csr_sic}, problem ${\text{P}_2^{CSR,low}}$ can be further re-expressed by the epigraph reformulation \cite{epigraph_reformulation}.
\vspace{-2pt}
\begin{equation}\nonumber
{\text{P}_{3}^{CSR,low}}:\mathop {\min }\limits_{\kappa,\varepsilon,\mathbf{\Theta}}\quad {{\frac{\kappa}{\varepsilon}}}
\end{equation}
\begin{equation}\nonumber
\text{s.t.}\quad\eqref{C_PSR_diag}, \eqref{39}
\end{equation}
\begin{equation}\label{CSR_2_1_a}
     \kappa \geq \sqrt{\sigma_0^4\left( 1+\gamma_{SIC} \right)-4 \sigma_0^2 \hat{p} \left|{ h_{SR}({\bf\Theta})}\right|^2}-\sigma_0^2+3\hat{p}\left|{ h_{SR}({\bf\Theta})}\right|^2,
\end{equation}
\begin{equation}\label{CSR_2_1_b}
    \kappa \geq \frac{\left(2^{\eta \epsilon_c}-1\right)\sigma_0^2}{\eta },
\end{equation}
\begin{equation}\label{CSR_2_1_c}
    0 \leq \varepsilon \leq \hat{p} \left|{ h_{SB}({\bf\Theta})}\right|^2 \left|{ h_{BR}({\bf\Theta})}\right|^2,
\end{equation}
\begin{equation}\label{CSR_2_1_d}
    0<\frac{\kappa}{\varepsilon}\leq 1,
\end{equation}
where $\kappa$ and $\varepsilon$ are introduced non-negative slack variables. 
According to \cite{epigraph_reformulation}, the optimal $\bf \Theta$ of ${\text{P}_{3}^{CSR,low}}$ is equivalent to the optimal $\bf \Theta$ of ${\text{P}_{2}^{CSR,low}}$, thereby implying that the solution of ${\text{P}_{2}^{CSR,low}}$ can be obtained by solving ${\text{P}_{3}^{CSR,low}}$.
\begin{itemize}
    \item High $\text{SNR}_p$ regime, i.e., $\hat{p} \left|{ h_{SR}({\bf\Theta})}\right|^2/{\sigma_0^2}\geq\frac{\gamma_{SIC}+1}{4}$. Note that under this condition, the constraint \eqref{R_s_csr} is always satisfied, thus the problem ${\text{P}_1^{CSR}}$ can be reduced to a low-complexity fraction-eliminated BPLM problem given by
\end{itemize}
\begin{equation}\nonumber
{\text{P}_2^{CSR,high}}:\mathop {\min }\limits_{\mathbf{\Theta}}\quad {\frac{\left(2^{\eta \epsilon_c}-1\right)\sigma_0^2}{\hat{p}\eta  \left|{ h_{SB}({\bf\Theta})}\right|^2 \left|{ h_{BR}({\bf\Theta})}\right|^2 }}
\end{equation}
\begin{equation}\nonumber
\text{s.t.}\quad\eqref{C_PSR_diag},
\end{equation}
\begin{equation}\label{CSR_P_3_a}
{\frac{\left(2^{\eta \epsilon_c}-1\right)\sigma_0^2}{\hat{p}\eta  \left|{ h_{SB}({\bf\Theta})}\right|^2 \left|{ h_{BR}({\bf\Theta})}\right|^2 }}\leq 1,
\end{equation}
\begin{equation}\label{CSR_P_3_b}
    \hat{p} \left|{ h_{SR}({\bf\Theta})}\right|^2/{\sigma_0^2}\geq\frac{\gamma_{SIC}+1}{4}.
\end{equation}

It is worth noting that the eliminated constraint \eqref{R_s_csr} can be reformulated to  \eqref{CSR_2_1_a} and \eqref{CSR_2_1_c} in the low $\text{SNR}_p$ regime. 
Considering a generic interior-point method for solving the above problems, ${\text{P}_2^{CSR,high}}$ exhibits a complexity reduction of $\mathcal{O}(2M^4)$ compared to ${\text{P}_3^{CSR,low}}$ \cite{complexity2}.

\subsection{The Optimization of IRS's Phase Shifts}

\subsubsection{The PSR case}
It can be observed that the minimization of ${\tilde{\alpha}_{lower}^{PSR}({\bf \Theta},\hat{p})}$ in $\text{P}_2^{PSR}$ can be equivalently transformed into the maximization of $\left|{ h_{SB}({\bf\Theta})}\right|^2 \left|{ h_{BR}({\bf\Theta})}\right|^2$. 
Therefore, we develop the PAP algorithm based on the Lipschitz descent lemma to maximize the squared modulus of the double reflection channel gain ${{\left|{ h_{SB}({\bf\Theta})}\right|^2 \left|{ h_{BR}({\bf\Theta})}\right|^2 }}$.
With some algebraic manipulations, $\text{P}_2^{PSR}$ can be equivalently re-written as a fraction-eliminated problem with a highly coupled ${\left|{ h_{SB}({\bf\Theta})}\right|^2 \left|{ h_{BR}({\bf\Theta})}\right|^2 }$.

\begin{equation}\nonumber
{\hat{\text{P}}_{2}^{PSR}}:\mathop {\max }\limits_{\mathbf{\Theta}}\quad {{\left|{ h_{SB}({\bf\Theta})}\right|^2 \left|{ h_{BR}({\bf\Theta})}\right|^2 }}
\end{equation}
\begin{equation}\nonumber
\text{s.t.}\quad\eqref{C_PSR_diag},
\end{equation}
\begin{equation}\label{P_2_1_Rs}
\sigma_0^2 (1+\gamma_c)\gamma_{SIC}  \leq \hat{p} \left| { h_{SR} ({\bf\Theta})} \right|^2,
\end{equation}
\begin{equation}\label{P_2_1_Rc}
    \sigma_0^2 \gamma_c \leq \hat{p} \left| { h_{SB} ({\bf\Theta})} \right|^2 \left| { h_{BR} ({\bf\Theta})} \right|^2.
\end{equation}

It is clear that $\hat{\text{P}}_{2}^{PSR}$ is a non-convex problem due to the highly coupled term $\left|{ h_{SB}({\bf\Theta})}\right|^2 \left|{ h_{BR}({\bf\Theta})}\right|^2$ in the objective function and the constraint \eqref{P_2_1_Rc}, which can be re-expressed as
\begin{equation}\label{Gamma_v}
\begin{array}{ll}
     &\left|{ h_{SB}({\bf\Theta})}\right|^2 \left|{ h_{BR}({\bf\Theta})}\right|^2=
    \left|{\bf g}_B {\bf\Theta} {\bf g}_S^H\right|^2 \left|{\bf g}_R {\bf\Theta} {\bf g}_B^H\right|^2  \\
     & = \left( {\boldsymbol v}^H {\bf G}_{SB} {\boldsymbol v}\right) \left( {\boldsymbol v}^H {\bf G}_{BR} \boldsymbol{v}\right) \triangleq \Gamma({\boldsymbol v}),
\end{array}
\end{equation}
where ${\boldsymbol v}=\text{diag}\left({\bf \Theta}\right)$ and ${{\bf G}_{ij}}=\text{diag}({\bf g}_i) {{\bf g}_j}^H \cdot \left(\text{diag}({\bf g}_i) {{\bf g}_j}^H\right)^H$ for ${i,j}\in \{{ S,B,R}\}$. 
Accordingly, we first approximate the lower bound of $\Gamma(\boldsymbol{v})$ with the following lemma.
\begin{Lemma}{\textit{(Lipschitz Descent Lemma \cite{Lip})}}
    Consider $\Gamma(\cdot)$ is a continuously differentiable function with Lipschitz continuous gradient and Lipschitz constant ${\mathcal L}$. Then, for all ${\boldsymbol v}_0 \in \mathbb{C}^{M \times 1}$, we have
    \begin{equation}
    \label{Lipschitz}
    \Gamma({\boldsymbol v}) \geq \Gamma({\boldsymbol v}_0)+\text{Re} \{ \nabla \Gamma({\boldsymbol v}_0)^H ({\boldsymbol v}-{\boldsymbol v}_0) \}-\frac{\mathcal L}{2}\left\| {\boldsymbol v}-{\boldsymbol v}_0 \right\|^2,    
\end{equation}
\end{Lemma}
where $\nabla \Gamma$ called the Lipschitz henceforth. 
Applying \eqref{Lipschitz} into \eqref{Gamma_v}, we further have the following approximate expression.
\begin{equation}\label{ap-Gamma_v}
    \Gamma(\boldsymbol{v}) \geq \frac{\mathcal L}{2}\text{Tr}\left({\bf U V}\right)+const,
\end{equation}
with ${\bf U}$, ${\bf V}$ and $const$ being defined as
\begin{equation}\label{int_U}
    {\bf U} \triangleq
    \begin{bmatrix}
    {\bf I}_{M} & -\frac{2}{\mathcal L}\mathcal{W}{\boldsymbol{v}}_0-{\bf I}_M{\boldsymbol{v}}_0 \\
    \left(-\frac{2}{\mathcal L}\mathcal{W}{\boldsymbol{v}}_0-{\bf I}_M{\boldsymbol{v}}_0\right)^H& 0 
    \end{bmatrix},
\end{equation}
\begin{equation}\label{int_V}
    {\bf V}\triangleq\begin{bmatrix}
    {\boldsymbol{v}} \\
    1
    \end{bmatrix}
    \cdot
    \begin{bmatrix}
    {\boldsymbol{v}} & 1
    \end{bmatrix},
\end{equation}
\begin{equation}\label{int_const}
    const \triangleq {\boldsymbol v}_0^H {\bf G}_{SB} {\boldsymbol v}_0 {\boldsymbol v}_0^H {\bf G}_{BR} {\boldsymbol v}_0 - 2 {\boldsymbol v}_0^H \mathcal{W} {\boldsymbol v}_0 - \frac{\mathcal L}{2}\left\| {\boldsymbol v}_0 \right\|^2, 
\end{equation}
where ${\mathcal{W}}={\bf G}_{SB}{\boldsymbol v}_0 {\boldsymbol v}_0^H {\bf G}_{BR} + {\bf G}_{BR} {\boldsymbol v}_0 {\boldsymbol v}_0^H {\bf G}_{SB}$. 
Therefore, the \eqref{P_2_1_Rc} can be approximated as
\begin{equation}\label{re-P_2_1_Rc}
    \frac{\sigma_0^2 \gamma_c}{\hat{p}} \leq  {\frac{\mathcal L}{2}\text{Tr}({\bf UV})}+const.
\end{equation}

Then, we tackle the non-convex term in \eqref{P_2_1_Rs}.
Let us denote ${{\bf Q}_{ij}} \triangleq \left[ {{\bf G}_{ij}},\Vec{{\bf 0}}^H; \Vec{{\bf 0}},0\right]$ for ${i,j}\in \{{ S,B,R}\}$. Due to the fact that $\left|{{ h}_{ij}({\bf\Theta})}\right|^2=\left|{{\bf g}_i {\bf \Theta g}_j}\right|^2=\boldsymbol{v}^H{{\bf G}_{ij}}\boldsymbol{v}=\text{Tr}\left({{\bf Q}_{ij}{\bf V}}\right)$, the constraint \eqref{P_2_1_Rs} can be rewritten as
\begin{equation}\label{re-P_2_1_Rs}
\sigma_0^2 (1+\gamma_c)\gamma_{SIC}  \leq \hat{p} \text{Tr}\left({\bf Q}_{SR}{\bf V}\right).
\end{equation}

Applying \eqref{ap-Gamma_v}, \eqref{re-P_2_1_Rc} and \eqref{re-P_2_1_Rs} into ${\hat{\text{P}}_{2}^{PSR}}$, problem ${\hat{\text{P}}_{2}^{PSR}}$ can be approximated as follows.
\begin{equation}\nonumber
{{\tilde{\text{P}}}_{2}^{PSR}}:\mathop {\max }\limits_{\mathbf{V}}\quad {\frac{\mathcal L}{2}\text{Tr}({\bf UV})}+const
\end{equation}
\begin{equation}\nonumber
\text{s.t.}\quad \eqref{re-P_2_1_Rc}, \, \eqref{re-P_2_1_Rs},
\end{equation}
\begin{equation}\label{rank1}
    {\bf V}\succeq 0,\,\text{rank}({\bf V})=1,\,{\bf V}_{\left(m,m\right)}=1,\,m=1,...,M+1.
\end{equation}

Note that ${{\tilde{\text{P}}}_{2}^{PSR}}$ is a semidefinite program with an {extra} rank-one constraint. Hence, the sequential rank-one constraint relaxation (SROCR) algorithm can be employed to obtain the rank-one {Karush-Kuhn-Tucker (KKT)} solution of ${{\tilde{\text{P}}}_{2}^{PSR}}$ \cite{SROCR}.

\begin{algorithm}[t]
  \caption{{The PAP Algorithm for PSR Case}}
  \label{alg1}
  \begin{algorithmic}[1]
    \STATE Initialize ${\boldsymbol v_0}^{\left(1\right)}$, error tolerance $\delta>0$, and set iteration index $l=1$. Then determine ${\bf U}^{\left(1\right)}$ and $const^{\left(1\right)}$ according to \eqref{int_U} and \eqref{int_const}, respectively;
    \REPEAT
    \STATE With ${\bf U}^{\left(l\right)}$ and $const^{\left(l\right)}$, solve ${{\tilde{\text{P}}}_{2}^{PSR}}$ through SROCR algorithm to get ${\bf V}^{\left(l+1\right)}$, with which ${\boldsymbol v_0}^{\left(l+1\right)}$ can be obtained by using eigenvalue decomposition;
    \STATE Update ${\bf U}^{\left(l\right)}\rightarrow{\bf U}^{\left(l+1\right)}$, $const^{\left(l\right)}\rightarrow const^{\left(l+1\right)}$ and $l\rightarrow l+1$;
    \UNTIL The increase of the objective value is below the threshold $\delta$.
  \end{algorithmic}
\end{algorithm}

\subsubsection{The CSR case}

Note that ${\text{P}_1^{CSR}}$ has been reformulated as a BPLM problem. To minimize the power leakage from the BD, we develop the PLM algorithm based on the successive convex approximation (SCA) method and the Lipschitz descent lemma under the following two $\text{SNR}_p$ regimes:

\begin{itemize}
    \item Low $\text{SNR}_p$ regime, i.e., $\hat{p} \left|{ h_{SR}({\bf\Theta})}\right|^2/{\sigma_0^2}<\frac{\gamma_{SIC}+1}{4}$.
\end{itemize}

It is noticeable that ${\text{P}_{3}^{CSR,low}}$ is also intractable due to the non-convex term in the objective function, \eqref{CSR_2_1_a} and \eqref{CSR_2_1_c}. 
To tackle the coupled term in ${\frac{\kappa}{\varepsilon}}$, the objective function of ${\text{P}_{3}^{CSR,low}}$ can be equivalently transformed to $\ln \kappa - \ln \epsilon$, and the SCA method based on the first-order Taylor series expansion is performed around $\kappa^{(l)}$ to get a safe linear approximation of the primitive
\begin{equation}
    \ln \kappa - \ln \epsilon \approx \ln \kappa^{(l)} + \frac{1}{\kappa^{(l)}}\left(\kappa-\kappa^{(l)}\right) - \ln \epsilon,
\end{equation}
where $\kappa^{(l)}$ is the $l$-th iteration point of the first-order Taylor series expansion.

Owing to the fact that $\left|{{ h}_{ij}({\bf\Theta})}\right|^2=\text{Tr}\left({\bf Q}_{ij}{\bf V}\right)$ for ${i,j}\in \{{ S,B,R}\}$, the constraints \eqref{39} and \eqref{CSR_2_1_a} can be rewritten as follows
\vspace{-2pt}
\begin{equation}\label{re-CSR_2_1_e}
    \sigma_0^2\left( 1+\gamma_{SIC} \right) > 4\hat{p} \text{Tr}\left({\bf Q}_{SR}{\bf V}\right),
\end{equation}
\begin{equation}\label{re-CSR_2_1_a}
    \kappa \geq \underbrace{\sqrt{\sigma_0^4\left(1+\gamma_{SIC} \right)-4 \sigma_0^2 \hat{p} \text{Tr}\left({\bf Q}_{SR}{\bf V}\right)}}_{\triangleq\chi\left(\text{Tr}\left({\bf Q}_{SR}{\bf V}\right)\right)}-\sigma_0^2+3\hat{p}\text{Tr}\left({\bf Q}_{SR}{\bf V}\right).
\end{equation}

Note that \eqref{re-CSR_2_1_a} is still hard to tackle due to the square root term in $\chi\left(\text{Tr}\left({\bf Q}_{SR}{\bf V}\right)\right)$. Applying the first-order Taylor series expansion, the upper bound of $\chi\left(\text{Tr}\left({\bf Q}_{SR}{\bf V}\right)\right)$ can be expressed as
\begin{equation}\small
\hspace{-10pt}\begin{array}{ll}\small
      &\chi\left(\text{Tr}\left({\bf Q}_{SR}{\bf V}\right)\right)  \\
      &\leq \underbrace{\chi\left(\text{Tr}\left({\bf Q}_{SR}{\bf V}^{(l)}\right)\right)-\frac{4 \sigma_0^2 \hat{p} \left[\text{Tr}\left({\bf Q}_{SR}{\bf V}\right)-\text{Tr}\left({\bf Q}_{SR}{\bf V}^{(l)}\right)\right]}{2\sqrt{\sigma_0^4\left(1+\gamma_{SIC} \right)-4 \sigma_0^2 \hat{p} \text{Tr}\left({\bf Q}_{SR}{\bf V}^{(l)}\right)}}}_{\triangleq \chi\left(\text{Tr}\left({\bf Q}_{SR}{\bf V}\right)\right)_{upper}},
\end{array}
\end{equation}
where ${\bf V}^{(l)}\in \mathbb{C}^{\left(M+1\right) \times \left(M+1\right)}$ is the $l$-th iteration point within the feasible region of ${\text{P}_{3}^{CSR,low}}$. Thus, the constraint \eqref{CSR_2_1_a} can be approximated as
\begin{equation}\label{re-CSR_2_1_a2}
    \kappa \geq \chi\left(\text{Tr}\left({\bf Q}_{SR}{\bf V}\right)\right)_{upper}-\sigma_0^2+3\hat{p}\text{Tr}\left({\bf Q}_{SR}{\bf V}\right).
\end{equation}

Subsequently, we proceed to address the non-convex term in constraint \eqref{CSR_2_1_c} with the help of the Lipschitz descent lemma. Following the rationale from \eqref{P_2_1_Rc} to \eqref{re-P_2_1_Rc}, \eqref{CSR_2_1_c} can be approximated as
\begin{equation}\label{re-CSR_2_1_c}
    \varepsilon \leq  {\frac{\mathcal L}{2}\text{Tr}({\bf UV})}+const.
\end{equation}

Applying \eqref{re-CSR_2_1_e}, \eqref{re-CSR_2_1_a2} and \eqref{re-CSR_2_1_c} to $\text{P}_{3}^{CSR,low}$, 
the approximation of ${\text{P}}_{3}^{CSR,low}$ can be expressed as follows.
\begin{equation}\nonumber
{{\tilde{\text{P}}}_{3}^{CSR,low}}:\mathop {\min }\limits_{\kappa,\varepsilon,\bf{V}}\quad {\ln \kappa^{(l)} + \frac{1}{\kappa^{(l)}}\left(\kappa-\kappa^{(l)}\right) - \ln \epsilon}
\end{equation}
\begin{equation}\nonumber
\text{s.t.}\quad\eqref{CSR_2_1_b},\,\eqref{CSR_2_1_d},\,\eqref{rank1},\,\eqref{re-CSR_2_1_e},\, \eqref{re-CSR_2_1_a2},\, \eqref{re-CSR_2_1_c}.
\end{equation}

Note that ${{\tilde{\text{P}}}_{3}^{CSR,low}}$ is also a semidefinite program with an {extra} rank-one constraint, which can be solved by the SROCR algorithm.
\begin{itemize}
    \item High $\text{SNR}_p$ regime, i.e., $\hat{p} \left|{ h_{SR}({\bf\Theta})}\right|^2/{\sigma_0^2}\geq\frac{\gamma_{SIC}+1}{4}$.
\end{itemize}

In the high $\text{SNR}_p$ regime, the constraint \eqref{R_s_csr} can be omitted, which reduces the complexity of the optimization problem.
Similar to the PSR case, the minimization of the objective function in ${\text{P}_2^{CSR,high}}$ can be equivalently transformed into the maximization of $\left|{ h_{SB}({\bf\Theta})}\right|^2 \left|{ h_{BR}({\bf\Theta})}\right|^2$. 

Note that ${\text{P}_2^{CSR,high}}$ is mathematically intractable due to the non-convexity of the objective function and the constraints \eqref{CSR_P_3_a}, \eqref{CSR_P_3_b}.
Following the rationale from \eqref{Gamma_v} to \eqref{ap-Gamma_v}, the objective function and \eqref{CSR_P_3_a} can be, respectively, rewritten as
\begin{equation}\label{ob}
    \mathop {\max }\limits_{\mathbf{V}}\quad {\frac{\mathcal L}{2}\text{Tr}({\bf UV})}+const,
\end{equation}
\begin{equation}\label{re-CSR_P_3_a}
   {\frac{\left(2^{\eta \epsilon_c}-1\right)\sigma_0^2}{\hat{p}\eta}}\leq {\frac{\mathcal L}{2}\text{Tr}({\bf UV})}+const.
\end{equation}

Utilizing a similar manipulation from \eqref{39} to \eqref{re-CSR_2_1_e}, constraint \eqref{CSR_P_3_b} can be expressed as
\begin{equation}\label{re-CSR_P_3_b}
     \sigma_0^2\left( 1+\gamma_{SIC} \right) \leq 4\hat{p} \text{Tr}\left({\bf Q}_{SR}{\bf V}\right).
\end{equation}

{Finally, applying \eqref{ob}, \eqref{re-CSR_P_3_a}, and \eqref{re-CSR_P_3_b} to ${\text{P}_2^{CSR,high}}$, the optimization problem in the CSR case at high $\text{SNR}_p$ can be formulated as a convex problem with an extra rank-one constraint.}
\begin{equation}\nonumber
{{\tilde{\text{P}}}_2^{CSR,high}}: \mathop {\max }\limits_{\mathbf{V}}\quad {\frac{\mathcal L}{2}\text{Tr}({\bf UV})}+const
\end{equation}
\begin{equation}\nonumber
\text{s.t.}\quad\eqref{rank1},\,\eqref{re-CSR_P_3_a},\,\eqref{re-CSR_P_3_b}.
\end{equation}

Note that ${{\tilde{\text{P}}}_2^{CSR,high}}$ and ${{\tilde{\text{P}}}_2^{PSR}}$ have a similar structure, thus the SROCR algorithm can be applied to generate a rank-one KKT solution with favorable complexity.

\begin{algorithm}[t]
  \caption{{The PLM Algorithm for CSR Case}}
  \label{alg2}
  \begin{algorithmic}[1]
    \STATE Initialize ${\boldsymbol v_0}^{\left(1\right)}$, ${\kappa}^{\left(1\right)}$, error tolerance $\delta>0$, and set iteration index $l=1$. Then determine ${\bf U}^{\left(1\right)}$, ${\bf V}^{(1)}$ and $const^{\left(1\right)}$ according to \eqref{int_U}, \eqref{int_V} and \eqref{int_const}, respectively;
    \IF{{$\hat{p} \left|{ h_{SR}({\bf\Theta})}\right|^2/{\sigma_0^2}<\frac{\gamma_{SIC}+1}{4}$}}
    \REPEAT
    \STATE Given ${\bf U}^{\left(l\right)}$, ${\bf V}^{\left(l\right)}$, $\kappa^{(l)}$ and $const^{\left(l\right)}$, solve ${\tilde{\text{P}}}_{3}^{CSR,low}$ through SROCR algorithm to get ${\bf V}^{\left(l+1\right)}$, ${\kappa}^{\left(l+1\right)}$ and ${\varepsilon}^{\left(l+1\right)}$, with which ${\boldsymbol v_0}^{\left(l+1\right)}$ can be obtained by using eigenvalue decomposition;
    \STATE Update ${\bf U}^{\left(l\right)}\rightarrow{\bf U}^{\left(l+1\right)}$, ${\bf V}^{\left(l\right)}\rightarrow{\bf V}^{\left(l+1\right)}$, $const^{\left(l\right)}\rightarrow const^{\left(l+1\right)}$ and $l\rightarrow l+1$;    
    \UNTIL The increase of the objective value is below the threshold $\delta$.
    \ELSE
    \REPEAT
    \STATE Given ${\bf U}^{\left(l\right)}$ and $const^{\left(l\right)}$, solve ${\tilde{\text{P}}}_{2}^{CSR,high}$ through SROCR algorithm to get ${\bf V}^{\left(l+1\right)}$, with which ${\boldsymbol v_0}^{\left(l+1\right)}$ can be obtained by using eigenvalue decomposition;
    \STATE Update ${\bf U}^{\left(l\right)}\rightarrow{\bf U}^{\left(l+1\right)}$, $const^{\left(l\right)}\rightarrow const^{\left(l+1\right)}$ and $l\rightarrow l+1$;    
    \UNTIL The increase of the objective value is below the threshold $\delta$.
    \ENDIF
  \end{algorithmic}
\end{algorithm}

\subsection{Overall procedure of the PAP and PLM algorithms}\label{sub3}

Under the statistical WCSI assumption, we summarized the developed PAP and PLM algorithms in Algorithms 1 and 2 for the PSR and CSR cases, respectively.
Therefore, the optimizations of ${\bf \Theta}$, $\alpha$, and $p$ are performed for the optimal W's detection threshold. Note that for the non-WCSI scenario, Algorithms 1 and 2 still hold with some minor modifications, which will be discussed as follows.

In the non-WCSI scenario, the legitimate system has no prior knowledge regarding the location or the detection threshold of W. 
To maximize the average DEP under the non-WCSI scenario, ${\text{P}_1^{PSR}}$ and ${\text{P}_1^{CSR}}$ can also be reformulated as a BPLM problem by employing the designed strategy presented in Proposition \ref{Pro2}. Consider that W adopts a fixed detection threshold in the non-WCSI scenario, the optimal $p$ can be determined using \eqref{fixed_p} for maximizing the average DEP. With the obtained $p$, the BPLM problem for the PSR and CSR cases can also be solved by the PAP and the PLM algorithm, respectively.
In the following, we analyze the convergence and complexity of the proposed algorithms.

{\bf Convergence:}
Regarding Algorithm 1, the objective function value remains monotonically non-decreasing at each step of the proposed algorithm, ${\frac{\mathcal L}{2}\text{Tr}({\bf U}^{(l+1)}{\bf V}^{(l+1)})}+const^{(l+1)}$ $\geq$ ${\frac{\mathcal L}{2}\text{Tr}({\bf U}^{(l)}{\bf V}^{(l)})}+const^{(l)}$ holds after steps 2-5 of Algorithm 1. Meanwhile, $\frac{\mathcal L}{2}\text{Tr}({\bf U}{\bf V})+const$ is continuous over the compact feasible set of ${{\tilde{\text{P}}}_{2}^{PSR}}$. Thus, the upper bound of the objective function can be reached within a finite number of iterations \cite{JMY}, which means that the proposed algorithm can eventually converge and terminate accordingly.

{\bf Complexity:}
For the optimization of ${{\tilde{\text{P}}}_{2}^{PSR}}$, the proposed algorithm using the interior-point method incurs complexity $C^{PSR}\triangleq\mathcal{O}\left[\ln\left(\frac{1}{\delta}\right)\sqrt{4M+5}\right]\{n\left[4(M+1)^3+1\right]+n^2\left[4(M+1)^2+1\right]+n^3\}=\mathcal{O}\{\ln\left(\frac{1}{\delta}\right)M^{6.5}\}$ \cite{complexity2}. {Similarly, the corresponding complexity of ${{\tilde{\text{P}}}_{3}^{CSR,low}}$ and ${{\tilde{\text{P}}}_{2}^{CSR,high}}$ can be expressed as $C^{CSR,low}=C^{CSR,high}\triangleq\mathcal{O}\{\ln\left(\frac{1}{\delta}\right)M^{6.5}\}$.}

\section{Simulation Results}

In this section, we present numerical results to evaluate the covert performance of the IRS-assisted SR communication system. Assume that S, BD, R, W and IRS are located at $(0,0)$, $(20,0)$, $(40,0)$ $(45,0)$ and $(20,25)$ in meter(m) in a two-dimensional plane, respectively. In this scenario, the non-line-of-sight (NLoS) component ${ h}_{NLoS}$ and LoS component ${ h}_{LoS}$ are considered between each node. Specifically, ${ h}_{NLoS}$ is modeled as Rayleigh fading, and the phase of ${ h}_{LoS}$ follows the uniform distribution over $2\pi$ radians while its amplitude remains unity. 
Considering the Rician fading and path loss attenuation, we have ${ h}=\tilde{ h}/{\sqrt{L(d)}}$, where $d$ denotes the distance and $\tilde{ h}\triangleq\left(\sqrt{\frac{\mathcal{B}}{1+\mathcal{B}}}{ h}_{LoS}+\sqrt{\frac{1}{1+\mathcal{B}}}{ h}_{NLoS}\right)$ for ${\tilde{ h}\in\{{\bf g}_S,{\bf g}_B,{\bf g}_R,{\bf g}_W\}}$.
The effective path loss function is modeled as $L(d)(\text{in dB})=35.1+36.7\lg(d)-G_t-G_r$, where $G_t$ and $G_r$ denote the transmitter and R antenna gains with $G_t=G_r= 10 \text{dBi}$ \cite{parameter1}. The other parameters are set as $\mathcal{B}=3$, $\sigma_0^2=-80\text{dBm}$, ${\mathcal L}=2.5\times10^{-3}$.
In this paper, we assume that $\eta$ is large enough, which allows $c(k)$ to be completely treated as a multi-path component in the CSR case. The simulation result under this assumption can be regarded as the performance upper bound of the practical CSR communication scenario.
Moreover, the Monte-Carlo simulations are averaged over $10^5$ independent trials.
To verify the superiority of the proposed algorithm, the optimal detection threshold is assumed in Fig. \ref{convergence}-\ref{N_b}.

\begin{figure}
\centering
\includegraphics[width=0.3\textwidth]{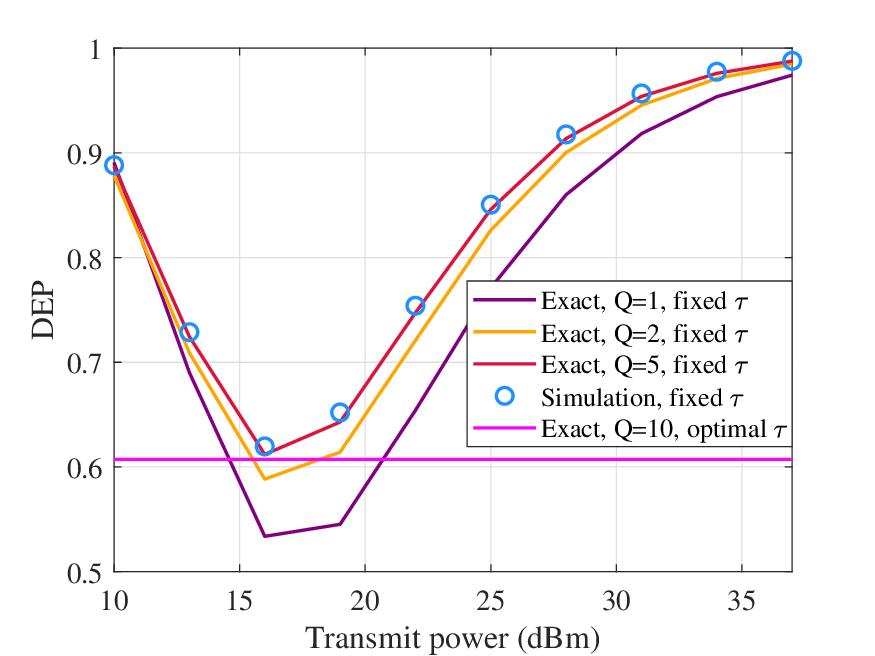}
\caption{{The average DEP versus the average transmit power of the source with $\alpha=0.2$.}}
\label{MD_FA_DEP}
\vspace{-10pt}
\end{figure}

Fig. \ref{MD_FA_DEP} shows the average DEP $\xi_{W,\tau}$ as a function of the transmit power $P_{\max}$.
It is observed that the exact result in \eqref{DEP_G_C} matches well with the simulated one for a relatively small complexity-accuracy tradeoff parameter $Q(Q=5)$, which verifies the accuracy of the derived analytical results.
Considering the availabilities of W's CSI, the average DEP curve shows different trends with the transmit power. Specifically, for the statistical WCSI scenario, the optimal detection threshold of W can be obtained, and the average DEP curve remains at a constant level.
This phenomenon reveals that increasing the power of the public signal does not always enhance the covert performance of the proposed SR communication system. This is because it also increases the leakage of covert signal power.
For the non-WCSI scenario, assume that W adopts a fixed detection threshold, and the average DEP possesses a local minimum point with respect to the transmit power. The simulation results in Fig. \ref{MD_FA_DEP} validate Proposition 3.

\begin{figure}
\centering
\includegraphics[width=0.3\textwidth]{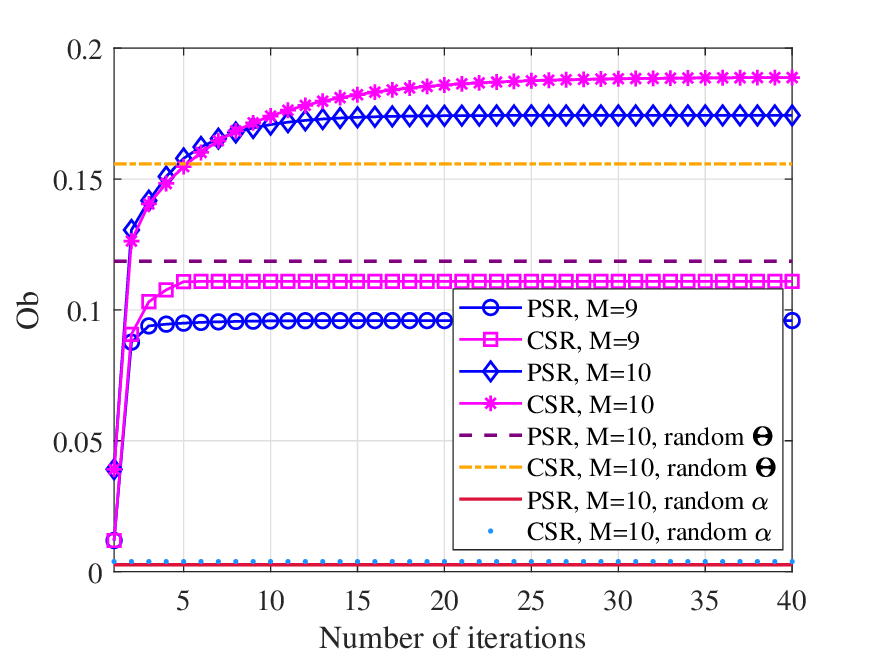}
\caption{{Convergence behavior of the developed algorithm with $P_{\max}=\text{25dBm}$, $\epsilon_{sic}=\text{2bps/Hz}$ and $\epsilon_{c}=\text{0.5bps/Hz}$.}}
\label{convergence}
\vspace{-10pt}
\end{figure}

Fig. \ref{convergence} illustrates the convergence of the proposed algorithm and the benchmark performance obtained from the random phase shifts of IRS and the random reflection coefficient of BD, where $\text{Ob}$ is the objective function of ${\tilde{\text{P}}}_{2}^{PSR}$, ${{\tilde{\text{P}}}_{3}^{CSR,low}}$ and ${{\tilde{\text{P}}}_{2}^{CSR,high}}$. As the number $M$ of phase shifters increases, the proposed algorithm achieves better covert performance, however, it requires more iterations to converge. 
Moreover, due to constraints \eqref{re-CSR_2_1_e} and \eqref{re-CSR_P_3_b}, the proposed algorithm requires more iterations to achieve convergence in the CSR case compared to the PSR case.

\begin{figure}
\centering
\includegraphics[width=0.3\textwidth]{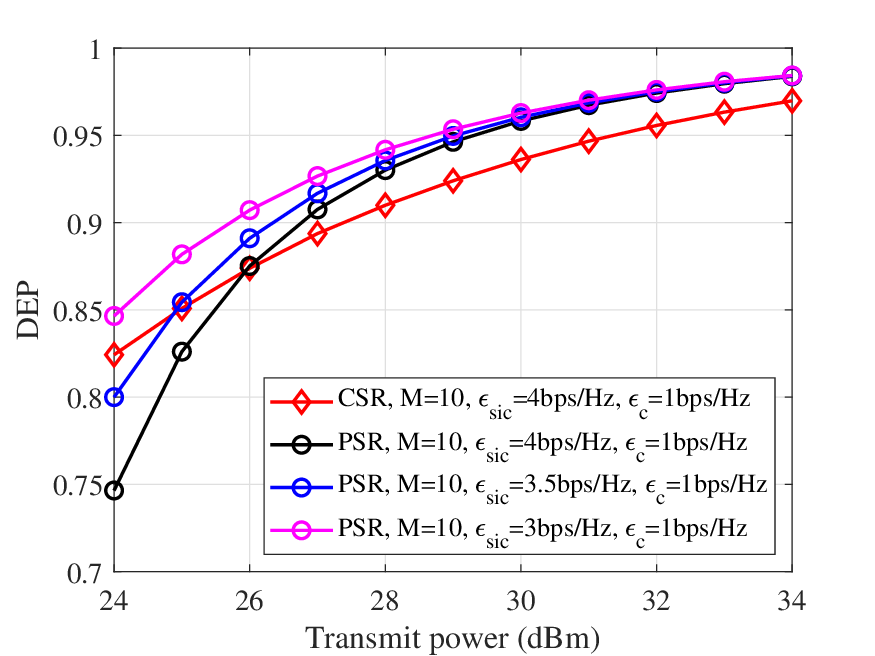}
\caption{{Maximal average DEP versus the transmit power of the source.}}
\label{change_SNR}
\vspace{-10pt}
\end{figure}

Fig. \ref{change_SNR} depicts the maximal average DEP versus the transmit power $P_{\max}$ for the proposed algorithm under the optimal detection threshold at W. 
It is worth noting that, unlike the results shown in Fig. \ref{MD_FA_DEP}, the average DEP increases with the transmit power. This is because employing more transmit power provides a higher degree of freedom to adjust the reflection coefficient of BD, which significantly improves covert performance. The simulation results also show that, at a low SNR level, the average DEP in the CSR case outperforms that in the PSR case, and as the primary rate requirement increases, the intersection of the PSR and CSR curves moves towards the right. This behavior implies that the performance of the PSR case is limited by the transmit power under the high primary rate condition.

\begin{figure}
\centering
\includegraphics[width=0.3\textwidth]{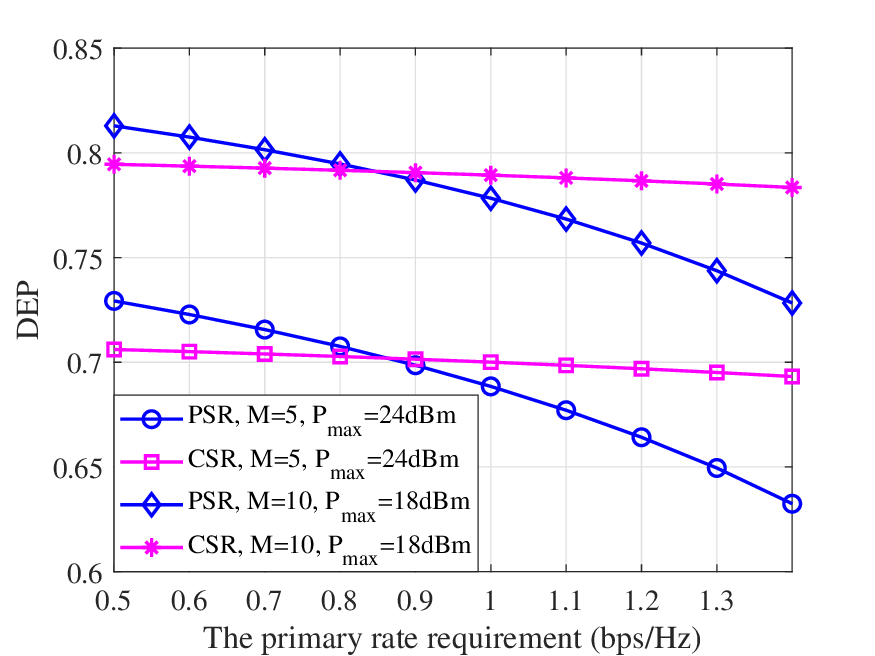}
\caption{{Maximal average DEP versus the primary rate requirement with $\epsilon_{c}=\text{0.5bps/Hz}$.}}
\label{epsilon_SIC}
\vspace{-10pt}
\end{figure}

\begin{figure}
\centering
\includegraphics[width=0.3\textwidth]{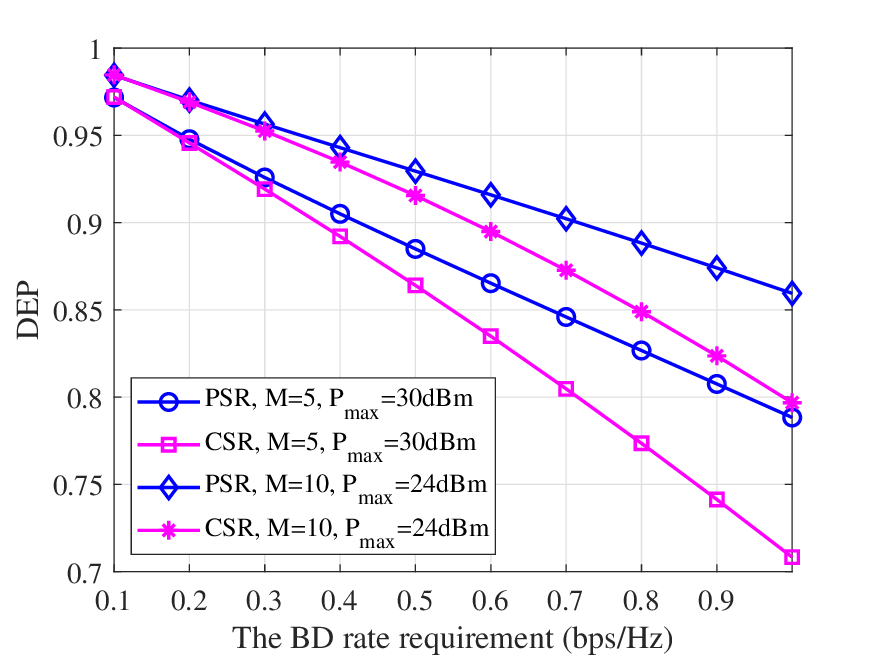}
\caption{{Maximal average DEP versus the BD rate requirement with $\epsilon_{SIC}=\text{2bps/Hz}$.}}
\label{epsilon_C}
\vspace{-10pt}
\end{figure}

Fig. \ref{epsilon_SIC} and \ref{epsilon_C} show the maximal average DEP versus the primary and the BD rate requirement, respectively. It is clear that increasing the number of IRS reflecting elements can save transmit power without compromising covert performance. 
{This is because the introduction of the passive IRS creates a double-reflection channel for the signal detection of BD.} Therefore, an increase of $M$ facilitates the generation of a large IRS phase-shift uncertainty to confuse W.
Furthermore, as shown in Fig. \ref{epsilon_SIC}, when the primary rate requirement increases, the average DEP in the PSR case decreases more compared to the CSR case, which means that improving the primary rate requirement primarily compromises the covert performance of the PSR case. However, in Fig. \ref{epsilon_C}, the slope of the average DEP curve in the CSR case is greater than that in the PSR case, indicating that a higher BD rate requirement has a greater limitation on the covert performance of the CSR case.  
Overall, the analysis of Fig. \ref{epsilon_SIC} and Fig. \ref{epsilon_C} provides a transmission strategy selection scheme for different covert SR communication scenarios. Specifically, the CSR strategy is more suitable for the scenario where the transmission rate requirement of the primary system is significantly higher than the BD system. On the contrary, when the BD system has a certain transmission rate requirement, the performance of the PSR strategy is superior.

\begin{figure}
\centering
\includegraphics[width=0.3\textwidth]{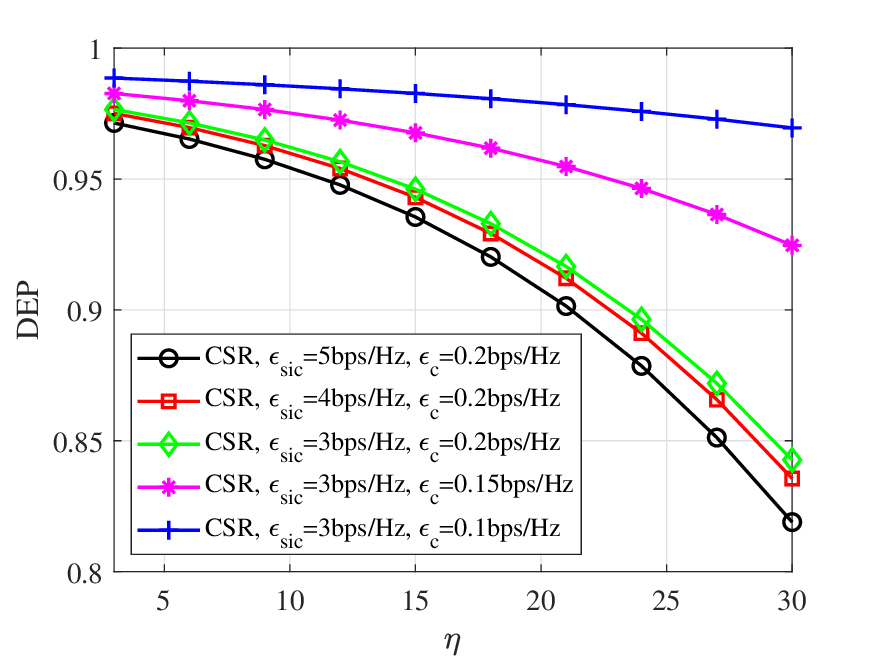}
\caption{{Maximal average DEP versus the ratio of the symbol period with $M=10$ and $P_{\max}=\text{25dBm}$.}}
\label{N_b}
\vspace{-10pt}
\end{figure}

Fig. \ref{N_b} plots the average DEP achieved by different primary/BD rate requirements versus the ratio of the symbol period under the assumption that the BD signal can be perfectly regarded as a multi-path component. 
{It can be observed that the average DEP decreases with the ratio of the symbol period, attributed to the fact shown in \eqref{CSR_rate_c} that a large $\eta$ is harmful to achieving a high BD rate requirement.}
However, in the practical CSR communication scenario, the channel capacity of the primary signal can be improved only when $\eta$ is large enough since the BD signal can be regarded as a multi-path component rather than interference.
Therefore, the selection of $\eta$ needs to consider the trade-off between the rate requirements of the primary and the BD signal. 
Further, as the BD rate requirement increases, the slope of the average DEP curve also increases. However, when the primary rate requirement increases, the slope of the average DEP curve remains nearly unchanged. This behavior implies that covert communication tends to prefer the CSR mode when the QoS of the primary signal is high.

\vspace{-10pt}
\section{Conclusion}
In this paper, we proposed an IRS-assisted SR communication scheme for enhancing the covertness of the BD signal transmission. We analyzed the average DEP and developed an optimal strategy to improve covert performance under the different WCSI scenarios. Aiming at the average DEP maximization, we designed the PAP and PLM algorithms for the PSR and CSR cases. Simulation results revealed that the PSR and CSR cases are suitable for the scenario with higher requirements of the BD rate and primary rate, respectively. In future work, we will investigate covert communication with IRS-assisted multi-antenna nodes and consider the impact of the direct links between each node. 
\vspace{-10pt}
\begin{appendices}
    \section{Proof of Proposition 1}\label{App_A}
    According to \cite{tight_M}, the two random variables are i.i.d. if they follow a joint Gaussian distribution and their cross-correlation is zero. Since ${\bf g}_S^H {\bf \Theta g}_B$, ${\bf g}_B^H {\bf \Theta g}_W \sim \mathcal{CN}\left(0,M\right)$, for $M\to \infty$. In the following, we first demonstrate the joint Gaussian distribution of ${\bf g}_S^H {\bf \Theta g}_B$ and ${\bf g}_B^H {\bf \Theta g}_W$.
    
    Denote ${\bf g}_i=\left[g_{i,1},...,g_{i,M}\right]^T$, where $i\in \{{B,S,W}\}$. Without loss of generality, the complex terms $g_{{ B},m}$, $g_{{ S},m}$ and $g_{{ W},m}$ can be defined as $g_{{ B},m}=b_m+jc_m$, $g_{{ S},m}=s_m+jt_m$ and $g_{{ W},m}=w_m+jx_m$, respectively, where $b_m$, $c_m$, $s_m$, $t_m$, $w_m$ and $x_m$ are i.i.d.
    Therefore, ${\bf g}_S^H {\bf \Theta g}_B$ can be re-expressed as
    \begin{equation}
    \begin{array}{ll}\label{65}
         & {\bf g}_S^H {\bf \Theta g}_B=\sum\limits_{m=1}^{M} e^{-j\theta_m} g_{{B},m} g_{{S},m}^\ast  \\
         & = \sum\limits_{m=1}^{M} \left(b_m s_m + c_m t_m\right)\cos\theta + \left(c_m s_m - b_m t_m\right)\sin\theta \\
         & \quad\quad + j\left[\left(c_m s_m - b_m t_m\right)\cos\theta - \left(b_m s_m + c_m t_m\right)\sin\theta\right]\\
         & = \sum\limits_{m=1}^{M} \left(T_m+jS_m\right) \left(c_m-jb_m\right),
    \end{array}
    \end{equation}
    with $S_m$ and $T_m$ being defined as
    \begin{equation}
    \begin{array}{ll}
         & S_m = s_m \cos\theta - t_m \sin \theta,  \\
         & T_m = t_m \cos\theta + s_m \sin \theta.
    \end{array}
    \end{equation}

    Following the same rationale, we can also re-writte ${\bf g}_B^H {\bf \Theta g}_W$ as
    \begin{equation}\label{67}
        {\bf g}_B^H {\bf \Theta g}_W = \sum\limits_{m=1}^{M} \left(W_m+jX_m\right) \left(b_m-jc_m\right),
    \end{equation}
    with $W_m$ and $X_m$ being given by
    \begin{equation}
    \begin{array}{ll}
         & W_m = w_m \cos\theta + x_m \sin \theta,  \\
         & X_m = x_m \cos\theta - w_m \sin \theta.
    \end{array}
    \end{equation}

    To prove the jointly Gaussian distribution of ${\bf g}_S^H {\bf \Theta g}_B$ and ${\bf g}_B^H {\bf \Theta g}_W$, we form a linear combination of them using $\beta_1$ and $\beta_2$ by applying the result in \eqref{65} and \eqref{67}.
    \begin{equation}
    \begin{array}{ll}
    & \beta_1{\bf g}_S^H {\bf \Theta g}_B + \beta_2{\bf g}_B^H {\bf \Theta g}_W  \\
    & = \sum\limits_{m=1}^{M} \beta_1 \left(T_m+jS_m\right) \left(c_m-jb_m\right) \\
    & \quad\quad + \beta_2 \left(W_m+jX_m\right) \left(b_m-jc_m\right)\\
    & = \sum\limits_{m=1}^{M} \left[\beta_1 \left(T_m+jS_m\right) -j\beta_2 \left(W_m+jX_m\right)\right]c_m\\ 
    & \quad\quad - \left[j\beta_1 \left(T_m+jS_m\right) - \beta_2 \left(W_m+jX_m\right)\right]b_m.
    \end{array}
    \end{equation}    
Note that $\left[\beta_1 \left(T_m+jS_m\right) -j\beta_2 \left(W_m+jX_m\right)\right]$ and $\left[j\beta_1 \left(T_m+jS_m\right) -\beta_2 \left(W_m+jX_m\right)\right]$ are composed of $\left(T_m+jS_m\right)$ and $\left(W_m+jX_m\right)$ with orthogonal $\beta_i$ and $j\beta_i$, $i\in\{1,2\}$, indicating that they are independent and identically Gaussian distributed.
Hence, the Gaussian distribution of the linear combination can be proved \cite{tight_M}, which implies the jointly Gaussian distribution of ${\bf g}_S^H {\bf \Theta g}_B$ and ${\bf g}_B^H {\bf \Theta g}_W$.

    Applying the result of \eqref{65} and \eqref{67}, the correlation between ${\bf g}_S^H {\bf \Theta g}_B$ and ${\bf g}_B^H {\bf \Theta g}_W$ is given by 
    \begin{equation}
    \begin{array}{ll}
         & \mathcal{E} \{\left({\bf g}_S^H {\bf \Theta g}_B\right)\left({\bf g}_B^H {\bf \Theta g}_W\right)\}\\
         &  = \mathcal{E} \{ \sum\limits_{m=1}^{M} \left(T_m+jS_m\right) \left(c_m-jb_m\right) \\
         & \quad\quad  \times\sum\limits_{m=1}^{M} \left(W_m+jX_m\right) \left(b_m-jc_m\right) \}=0,
    \end{array}
    \end{equation}
    which follows from the fact that $b_m$, $c_m$, $S_m$, $T_m$, $W_m$ and $X_m$ are i.i.d. with zero mean. Therefore, we have demonstrated the i.i.d. property of ${\bf g}_S^H {\bf \Theta g}_B$ and ${\bf g}_B^H {\bf \Theta g}_W$.
    
    \section{Proof of Theorem 1}\label{App_B}
    With the result in \eqref{DEP}, we derive the first derivative of $\xi_{W,\tau}$ with respect to $\tau(>\sigma_0^2)$ as
    \begin{equation}
    \begin{array}{ll}
         &\frac{\partial \xi_{W,\tau}}{\partial\tau}=\lambda l_1 e^{\frac{-\lambda l_1 (\tau-\sigma_0^2)}{p}} ( 2\lambda^2 \int_{0}^{\frac{l_2 (\tau-\sigma_0^2)}{\alpha p}} e^{\frac{l_1 \alpha \lambda x}{l_2}} \\
         & \quad \quad \quad \quad  \times K_0\left(2 \lambda \sqrt{x} \right) {\text{dx}} -1 ).
    \end{array}
    \end{equation}  
    By setting $\frac{\partial \xi_{W,\tau}}{\partial\tau}=0$, the optimal solution of $\tau^\star$ can be obtained by the numerical integration as
    \begin{equation}
        \int_{0}^{\frac{l_2 (\tau-\sigma_0^2)}{\alpha p}} e^{\frac{l_1 \alpha \lambda x}{l_2}} K_0\left(2 \lambda \sqrt{x} \right) {\text{dx}} = \frac{1}{2\lambda^2}.
    \end{equation}
    Thus, we prove Theorem 1.

    \section{Proof of Proposition 2}\label{App_C}
    According to \eqref{Pro_FA} and \eqref{Pro_MD}, the average DEP can be written as
    \begin{equation}\small
    \begin{array}{ll}\label{DEP2}
         &\xi_{W,\tau}  \\
         & = \left\{
         \begin{array}{ll}
              \quad 1, & \tau \leq \sigma_0^2,  \\
              \begin{array}{ll}
                \Pr\left(pX+\sigma_0^2 > \tau \right)  \\
                \quad \quad +\Pr\left(pX+\alpha pY+\sigma_0^2 < \tau \right), 
              \end{array}
                & \tau > \sigma_0^2,
         \end{array}
         \right.
    \end{array}
    \end{equation}
    where $X=\frac{|{\bf g}_S^H {\bf \Theta g}_W|^2}{l_1}$ and $Y=\frac{|{\bf g}_S^H {\bf \Theta g}_B \cdot {\bf g}_B^H {\bf \Theta g}_W|^2}{l_1 l_2}$. 
    
    Consider $\tau>\sigma_0^2$, we re-express \eqref{DEP2} by employing the law of total probability (LTP) as follows.
    \begin{equation}\label{DEP3}
        \begin{array}{ll}
             & \xi_{W,\tau}=\Pr\left(pX+\sigma_0^2 > \tau \right) +\Pr\left(pX+\alpha pY+\sigma_0^2 < \tau \right)  \\
             &  \quad= 1-\Pr\left(pX+\sigma_0^2 < \tau \right) +\Pr\left(pX+\alpha pY+\sigma_0^2 < \tau \right)  \\
             &  \quad\overset{\text{LTP}}{=}1-\Pr\left(pX+\sigma_0^2 < \tau, pX+\alpha pY+\sigma_0^2 < \tau \right)\\
             &  \quad\quad \quad  - \Pr\left(pX+\sigma_0^2 < \tau, pX+\alpha pY+\sigma_0^2 > \tau \right)\\
             &  \quad\quad \quad +\Pr\left(pX+\alpha pY+\sigma_0^2 < \tau \right)\\
             &  \quad= 1-\Pr\left(X<\frac{\tau-\sigma_0^2}{p}<X+\alpha Y\right).             
        \end{array}
    \end{equation}

    It is clear that $\xi_{W,\tau}$ is a monotonically decreasing function with respect to $\alpha$. Hence, the proof is complete.

    \begin{figure}
    \centering
    \includegraphics[width=0.3\textwidth]{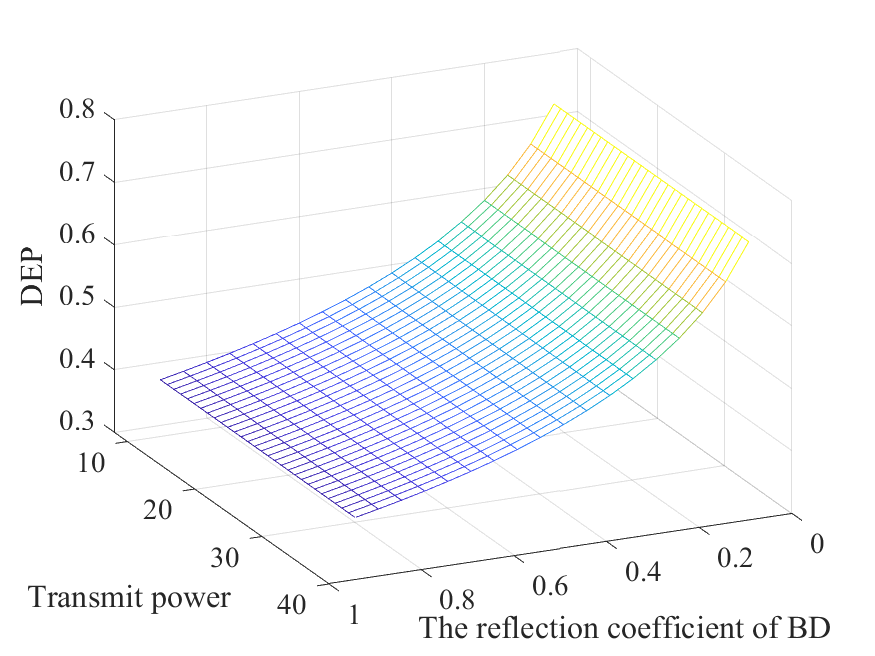}
    \caption{{The average DEP versus the transmit power of source and the reflection coefficient of BD with the optimal detection threshold.}}
    \label{optimal_tau2}
    \vspace{-10pt}
    \end{figure}
    
    \begin{figure}
    \centering
    \includegraphics[width=0.3\textwidth]{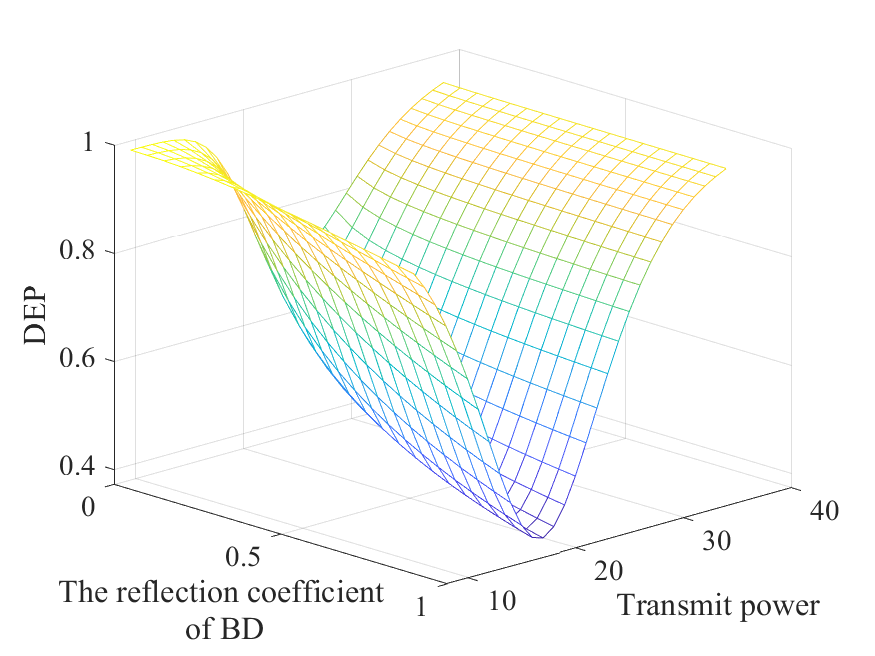}
    \caption{{The average DEP versus the transmit power of source and the reflection coefficient of BD with the fixed detection threshold.}}
    \label{fixed_tau}
    \vspace{-10pt}
    \end{figure}

    \section{Proof of Proposition 3}\label{App_D}

    The proposition can be proved under two different detection strategies as shown in the following two subsections.
    
    \subsection{For the optimal detection threshold of warden}
    We first derive the optimal detection threshold $\tau(>\sigma_0^2)$ with the result in \eqref{DEP3}.
    Denote $\omega=\frac{Y}{X}$, then \eqref{DEP3} can be re-written as
    \begin{equation}
        \xi_{W,\tau} = 1-\Pr\left(\frac{\tau-\sigma_0^2}{p\left[1+\alpha\omega(X)\right]}<X<\frac{\tau-\sigma_0^2}{p}\right).
    \end{equation}
    
    Note that for $M\to\infty$, $X$ follows an exponential distribution with parameter $\lambda l_1$. Thus, $\xi_{W,\tau}$ can be obtained as follows:
    \begin{equation}\label{76}
        \xi_{W,\tau} = 1-e^{\frac{-\lambda l_1(\tau-\sigma_0^2)}{p\left[1+\alpha\omega(X)\right]}}+e^{\frac{-\lambda l_1(\tau-\sigma_0^2)}{p}}.
    \end{equation}
    
    Subsequently, we derive the first derivative of $\xi_{W,\tau}$ with respect to $\tau$ as
    \begin{equation}
        \frac{\partial \xi_{W,\tau}}{\partial\tau} = \frac{\lambda l_1}{{p\left[1+\alpha\omega(X)\right]}}e^{\frac{-\lambda l_1(\tau-\sigma_0^2)}{p\left[1+\alpha\omega(X)\right]}} - \frac{\lambda l_1}{{p}}e^{\frac{-\lambda l_1(\tau-\sigma_0^2)}{p}}.
    \end{equation}
    By setting $\frac{\partial \xi_{W,\tau}}{\partial\tau}=0$, the optimal detection threshold is obtained as
    \begin{equation}\label{opt_tau}
        \tau^\star = \frac{p\left[1+\alpha\omega(X)\right]\ln\left[1+\alpha\omega(X)\right]}{\lambda l_1 \alpha\omega(X)}+\sigma_0^2.
    \end{equation}

    Applying \eqref{opt_tau} into \eqref{DEP3}, $\xi_{W,\tau^\star}$ can be re-expressed as
    \begin{equation}\small
    \label{79}
        \xi_{W,\tau^\star} = 1-\Pr\left(X<\frac{\left[1+\alpha\omega(X)\right]\ln\left[1+\alpha\omega(X)\right]}{\lambda l_1 \alpha \omega(X)}<X+\alpha Y\right).
    \end{equation}

    It is worth noting that the probability in \eqref{79} is independent of $p$. Therefore, the value of $p$ has no impact on the average DEP when W employs the optimal detection threshold. 
    
    \subsection{For the arbitrary and fixed detection threshold of warden}

    Considering the fixed detection threshold $\tau(>\sigma_0^2)$ of W, we derive the first derivative of $\xi_{W,\tau}$ with respect to $p$ by following the result in \eqref{76}.
    \begin{equation}
        \frac{\partial \xi_{W,\tau}}{\partial p} = \frac{\lambda l_1(\tau-\sigma_0^2)}{p^2} \left(e^{\frac{-\lambda l_1(\tau-\sigma_0^2)}{p}}-\frac{e^{\frac{-\lambda l_1(\tau-\sigma_0^2)}{p\left[1+\alpha\omega(X)\right]}}}{1+\alpha\omega(X)}\right).
    \end{equation}
    
    Let $\frac{\partial \xi_{W,\tau}}{\partial p}=0$, the point of the minimum value for $\xi_{W,\tau}$ can be represented as
    \begin{equation}\label{opt_p}
        p^\star = \frac{\lambda l_1 (\tau-\sigma_0^2)\omega(X)}{\left[1+\alpha \omega(X)\right]\ln\left[1+\alpha \omega(X)\right]}.
    \end{equation}

    Assume that the minimum point of $p$ satisfying the given constraint is $p_{\min}^{f}(<P_{\max})$, then $\xi_{W,\tau}$ has monotonicity in the following two regions by using the result in \eqref{opt_p}.
    \begin{itemize}
    \item If $0<p<p^\star$, $\xi_{W,\tau}$ is a monotonically decreasing function with respect to $p$, the optimal solution that maximizes $\xi_{W,\tau}$ is $p_{\min}^{f}$.
    \end{itemize}
    \begin{itemize}
    \item If $p^\star<p\leq P_{\max}$, $\xi_{W,\tau}$ is a monotonically increasing function with $p$, the optimal point of $p$ is $P_{\max}$.
    \end{itemize}

    Therefore, If $p$ has a solution within $\left[0, P_{\max}\right]$, the optimal $p$ can be obtained at the boundary of its feasible region as follows:
    \begin{equation}\label{fixed_p}
        p_{opt}=\arg\mathop{\max}\limits_{p_{\min}^f,P_{\max}} {\xi_{W,\tau}(p)}.
    \end{equation}

    The simulation results of the optimal and the fixed threshold for average DEP are presented in Fig. \ref{optimal_tau2} and \ref{fixed_tau}, respectively, which validate the correctness of our derivations.

\end{appendices}

\vfill


\begin{thebibliography}{99}

\bibitem{6G_0}
L. Lv \textit{et al}., “Safeguarding next-generation multiple access using physical layer security techniques: A tutorial,” \textit{Proc. IEEE}, early access, Mar. 25, 2024.

\bibitem{6G_1}
X.-H. You \textit{et al}., “Towards 6G wireless communication networks: Vision, enabling technologies, and new paradigm shifts,” \textit{Sci. China Inf. Sci.}, vol. 64, no. 1, pp. 1–74, Jan. 2021.

\bibitem{6G_2}
X.-H. You \textit{et al}., “Towards 6G TKµ extreme connectivity: Architecture, key technologies, and experiments,” \textit{IEEE Wireless Commun.}, vol. 30, no. 3, pp. 86–95, Jun. 2023.

\bibitem{6G_3}
Z. Zhang \textit{et al}., “6G wireless networks: Vision, requirements, architecture, and key technologies,” \textit{IEEE Veh. Technol. Mag.}, vol. 14, no. 3, pp. 28-41, Sep. 2019.

\bibitem{SR}
R. Long, Y. -C. Liang, H. Guo, G. Yang and R. Zhang, “Symbiotic radio: A new communication paradigm for passive internet of things,” \textit{IEEE Internet Things J.}, vol. 7, no. 2, pp. 1350-1363, Feb. 2020.

\bibitem{BD_1}
C. Xu \textit{et al}., “Practical backscatter communication systems for battery-free Internet of Things: A tutorial and survey of recent research,” \textit{IEEE Signal Process. Mag.}, vol. 35, no. 5, pp. 16–27,
2018.

\bibitem{BD_2}
Y. Ye, L. Shi, X. Chu, and G. Lu, “On the outage performance of ambient backscatter communications,” \textit{IEEE Internet Things J.}, vol. 7, no. 8, pp. 7265–7278, 2020.

\bibitem{BD_3}
Z. Ding, “Harvesting devices heterogeneous energy profiles and QoS requirements in IoT: WPT-NOMA vs BAC-NOMA,” \textit{IEEE Trans. Commun.}, vol. 69, no. 5, pp. 2837–2850, 2021

\bibitem{financial}
P. Zhang, M. Rostami, P. Hu, and D. Ganesan, “Enabling practical backscatter communication for on-body sensors,” in \textit{Proc. ACM SIGCOMM Conf.}, pp. 370–383, 2016.

\bibitem{health}
Z. Yang, Q. Huang, and Q. Zhang, “Nicscatter: Backscatter as a covert channel in mobile devices,” in \textit{Proc. Annu. Int. Conf. Mobile Comput. Net.}, pp. 356–367, 2017.

\bibitem{identity}
P. Wang, Z. Yan and K. Zeng, “BCAuth: Physical layer enhanced authentication and attack tracing for backscatter communications,” \textit{IEEE Trans. Inf. Forensics Security}, vol. 17, pp. 2818-2834, 2022.

\bibitem{LPD}
S. Yan, X. Zhou, J. Hu, and S. V. Hanly, “Low probability of detection communication: Opportunities and challenges,” \textit{IEEE Wireless Commun.}, vol. 26, no. 5, pp. 19–25, Oct. 2019

\bibitem{identity2}
H. Park and W. Lee, “Rethinking tag collisions for replay and relay attack resistance in backscatter networks,” \textit{IEEE Access}, vol. 11, pp. 31912-31923, 2023.

\bibitem{identity3}
H. Park, J. Yu, H. Roh and W. Lee, “SCBF: Exploiting a collision for authentication in backscatter networks,”  \textit{IEEE Commun. Lett.}, vol. 21, no. 6, pp. 1413-1416, June 2017.

\bibitem{covert_BD1}
W. Chen, H. Ding, S. Wang and F. Gong, “On the limits of covert ambient backscatter communications,” \textit{IEEE Trans. Wireless Commun.}, vol. 11, no. 2, pp. 308-312, Feb. 2022.

\bibitem{covert_BD2}
J. Liu, J. Yu, D. Niyato, R. Zhang, X. Gao and J. An, “Covert ambient backscatter communications with multi-antenna tag,” \textit{IEEE Trans. Wireless Commun.}, vol. 22, no. 9, pp. 6199-6212, Sept. 2023.

\bibitem{covert_BD3}
W. Ma, Z. Niu, W. Wang, S. He and T. Jiang, “Covert communication with uninformed backscatters in hybrid active/passive wireless networks: Modeling and performance analysis,”  \textit{IEEE Trans. Commun.}, vol. 70, no. 4, pp. 2622-2634, April 2022.

\bibitem{AN_source}
B. Gu, H. Xie and D. Li, “Act before another is aware: Safeguarding backscatter systems with covert communications,” \textit{IEEE Wireless Commun. Lett.}, vol. 12, no. 6, pp. 1106-1110, June 2023.

\bibitem{AN_destination}
J. Liu \textit{et al}., “Covert communication in ambient backscatter systems with uncontrollable RF source,” \textit{IEEE Trans. Commun.}, vol. 70, no. 3, pp. 1971-1983, March 2022.

\bibitem{AN_jammer}
Y. Xie, T. -T. Chan, X. Zhang, P. Lai and H. Pan, “Reflection-optimized covert communication for jammer-aided ambient backscatter systems,” in \textit{Proc. IEEE Global Communications Conference}, Kuala Lumpur, Malaysia, 2023, pp. 4277-4282.

\bibitem{IRS_SR1}
S. Basharat, S. A. Hassan, A. Mahmood, Z. Ding and M. Gidlund, “Reconfigurable intelligent surface-assisted backscatter communication: A new frontier for enabling 6G IoT networks,” \textit{IEEE Wireless Commun.}, vol. 29, no. 6, pp. 96-103, December 2022.

\bibitem{IRS_SR2}
Y. -C. Liang \textit{et al}., “Backscatter communication assisted by reconfigurable Intelligent Surfaces,” in \textit{Proceedings of the IEEE}, vol. 110, no. 9, pp. 1339-1357, Sept. 2022.

\bibitem{IRS_SR3}
M. Wu, X. Lei, X. Zhou, X. Tang and O. A. Dobre, “RIS-assisted energy- and spectrum-efficient symbiotic transmission in NOMA systems,” \textit{IEEE Trans. Commun.}, vol. 71, no. 5, pp. 2801-2815, May 2023.

\bibitem{IRS_SR4}
C. Zhou et al., “Energy-efficient maximization for RIS-aided MISO symbiotic radio systems,” \textit{IEEE Trans. Veh. Technol.}, vol. 72, no. 10, pp. 13689-13694, Oct. 2023.

\bibitem{IRS_SR5}
R. Liu \textit{et al}., “RIS-empowered satellite-aerial-terrestrial networks with PD-NOMA,” \textit{ IEEE Commun. Surveys Tuts.}, doi: 10.1109/COMST.2024.

\bibitem{semi-passive0}
Y. Wu, S. Wang, J. Luo and W. Chen, “Passive covert communications based on reconfigurable intelligent surface,” \textit{IEEE Wireless Commun. Lett.}, vol. 11, no. 11, pp. 2445-2449, Nov. 2022.

\bibitem{semi-passive1}
Z. Xu \textit{et al}., “Covert communication for intelligent reflecting surface based symbiotic radio systems,” in \textit{Proc. IEEE/CIC Int. Conf. Commun. China (ICCC)}, Dalian, China, 2023, pp. 1-6.

\bibitem{semi-passive2}
X. He, H. Xu, J. Wang, W. Xie, X. Li and A. Nallanathan, “Joint active and passive beamforming in RIS-assisted covert symbiotic radio based on deep unfolding,” \textit{IEEE Trans. Veh. Technol.}, doi: 10.1109/TVT.2024.3393724.

\bibitem{YXH}
X. Yu, D. Xu, Y. Sun, D. W. K. Ng and R. Schober, “Robust and secure wireless communications via intelligent reflecting surfaces,” \textit{IEEE J. Sel. Areas Commun.}, vol. 38, no. 11, pp. 2637-2652, Nov. 2020.


\bibitem{IRS-2}
Q. Wu and R. Zhang, “Intelligent reflecting surface enhanced wireless network via joint active and passive beamforming,” \textit{IEEE Trans. Wireless Commun.}, vol. 18, no. 11, pp. 5394–5409, Nov. 2019.

\bibitem{IRS-3}
Z. Yang, P. Yue, S. Wang, G. Pan and J. An, “Energy-efficient optimization for RIS-aided MIMO covert communications,” \textit{IEEE Internet Things J.}, vol. 10, no. 21, pp. 18993-19003, 1 Nov.1, 2023.

\bibitem{blockfading}
C. Boyer and S. Roy, “Backscatter communication and rfid: Coding, energy, and mimo analysis,” \textit{IEEE Trans. Commun.}, vol. 62, no. 3, pp. 770–785, 2014

\bibitem{pilot}
Z. Wang, L. Liu, and S. Cui, “Channel estimation for intelligent reflecting surface assisted multiuser communications: Framework, algorithms, and analysis,” \textit{IEEE Trans. Wireless Commun.}, vol. 19, no. 10, pp. 6607– 6620, June 2020.

\bibitem{Shannon}
C. E. Shannon, “A mathematical theory of communication,” \textit{Bell Syst.
Tech. J.}, vol. 27, no. 3, pp. 379–423, 1948.

\bibitem{coherence}
Y. Liang and V. V. Veeravalli, “Capacity of noncoherent time-selective Rayleigh-fading channels,” \textit{IEEE Trans. Inf. Theory,} vol. 50, no. 12, pp. 3095–3110, Dec. 2004

\bibitem{Hypotheses}
E. Lehmann and J. Romano, \textit{Testing Statistical Hypotheses}, 3rd ed. New York, NY, USA: Springer, 2005.

\bibitem{tight_M}
Z. Ding \textit{et al}., “On the impact of phase shifting designs on IRS-NOMA,” \textit{IEEE Wireless Commun. Lett.}, vol. 9, no. 10, pp. 1596–1600, Oct. 2020.

\bibitem{exp_PDF}
Q. Zhang, L. Zhang, Y. -C. Liang and P. -Y. Kam, “Backscatter-NOMA: A symbiotic system of cellular and internet-of-things networks,” \textit{IEEE Access}, vol. 7, pp. 20000-20013, 2019.

\bibitem{G-C}
Y. Liu, Z. Ding, M. Elkashlan and J. Yuan, “Nonorthogonal multiple access in large-scale underlay cognitive radio networks,” \textit{IEEE Trans. Veh. Technol.}, vol. 65, no. 12, pp. 10152-10157, Dec. 2016.

\bibitem{parameter1}
L. Lv, Q. Wu, Z. Li, Z. Ding, N. Al-Dhahir and J. Chen, “Covert communication in intelligent reflecting surface-assisted NOMA systems: Design, analysis, and optimization,” \textit{IEEE Trans. Wireless Commun.}, vol. 21, no. 3, pp. 1735-1750, March 2022.

\bibitem{epigraph_reformulation}
S. P . Boyd and L. V andenberghe,
\textit{Convex Optimization.} Cambridge, U.K.: Cambridge Univ. Press, 2004.

\bibitem{complexity2}
K. -Y. Wang \textit{et al}., “Outage constrained robust transmit optimization for multiuser MISO downlinks: Tractable approximations by conic optimization,” \textit{IEEE Trans. Signal Process.}, vol. 62, no. 21, pp. 5690-5705, Nov.1, 2014.

\bibitem{Lip}
Y. Sun, P. Babu and D. P. Palomar, “Majorization-minimization algorithms in signal processing, communications, and machine learning,” \textit{IEEE Trans. Signal Process.}, vol. 65, no. 3, pp. 794-816, 1 Feb.1, 2017.

\bibitem{SROCR}
P. Cao \textit{et al}., “A sequential constraint relaxation algorithm for rank-one constrained problems,” 2017 25th European Signal Processing Conference (EUSIPCO), Kos, Greece, 2017, pp. 1060-1064.

\bibitem{JMY}
M. Ji \textit{et al}., “Secure NOMA systems with a dual-functional RIS: Simultaneous information relaying and jamming,” \textit{IEEE Trans. Commun.}, vol. 71, no. 11, pp. 6514-6528, Nov. 2023.

\textbf{}

\end{thebibliography}
\end{document}